\documentclass[12pt,reqno]{amsart} % leqno: eq-numbers on right side

\usepackage{a4}
\usepackage{amsmath}
\usepackage{amssymb}
\usepackage{graphicx}               % for graphics

\theoremstyle{plain}            % body italics
\newtheorem{thm}{Theorem}[section]
\newtheorem{lem}[thm]{Lemma}
\newtheorem{cor}[thm]{Corollary}

\theoremstyle{definition}       % body roman

\theoremstyle{remark}
\newtheorem{rem}[thm]{Remark}
\newtheorem{exmp}[thm]{Example}

\numberwithin{equation}{section}

\DeclareMathOperator{\dom}    {dom}
\DeclareMathOperator{\spec}   {spec}
\DeclareMathOperator{\vol}    {vol}
\DeclareMathOperator{\tr} {tr}

\newlength{\maxbreite}%
\newlength{\maxhoehe}%
\newlength{\maxtiefe}%

\newcommand{\stelldrueber}[3][0pt]{%  Vorbereitung f"ur Kreis "uber Symbol
  \settowidth{\maxbreite}{#3}%
  \settoheight{\maxhoehe}{#3}%
  \settodepth{\maxtiefe}{#2}%
  \addtolength{\maxhoehe}{\maxtiefe}%
  {\makebox[\maxbreite]{\raisebox{\maxhoehe}{\hspace{#1}#2}}%
  \makebox[0pt][r]{#3}}%
}

\newcommand{\overcirc}[1]       % Kreis "uber Symbol
{\stelldrueber[.45ex]{$\scriptscriptstyle\circ$}{${#1}$}}

\newcommand{\R}{\mathbb{R}} % symbol for real numbers
\newcommand{\C}{\mathbb{C}} % symbol for complex numbers
\newcommand{\N}{\mathbb{N}} % symbol for natural numbers
\newcommand{\Z}{\mathbb{Z}} % symbol for integers
\newcommand{\Sphere}{\mathbb{S}} % symbol for sphere
\newcommand{\Torus}{\mathbb{T}} % symbol for torus
\newcommand{\eps}{\varepsilon} % shortcut
\renewcommand{\phi}{\varphi}   % shortcut
\newcommand{\eu}{\mathrm e}  %Euler number
\newcommand{\im}{\mathrm i} % complex unit

\newcommand{\HS}{\mathcal H}                % symbol for Hilbert/Sobolev space

\newcommand{\Ci} [1]{C^\infty ({#1})}       % C^\infty(#1)-spaces
\newcommand{\Cci}[1]{C_{\mathrm c}^\infty ({#1})} % C_c^\infty(#1)-spaces
\newcommand{\Cont}[2][{}]{C^{#1}({#2})}      % space of cont (C^k) functions
\newcommand{\Lsqr}[1]{L_2({#1})}            % L_2(#1)-spaces
\newcommand{\Sob}[2][1]{\HS^{#1}({#2})}     % Sobolev spaces
\newcommand{\Sobn}[2][1]{{\overcirc {\mathcal H}{}^{#1}({#2})}}% Sob-Raum H_0
\newcommand{\norm}[2][{}]{\|{#2}\|_{#1}}    % norm
\newcommand{\normsqr}[2][{}]{\|{#2}\|^2_{#1}} % norm squared
\newcommand{\bignormsqr}[2][{}]{\bigl\|{#2}\bigr\|^2_{#1}}% norm squared
\newcommand{\Bignormsqr}[2][{}]{\Bigl\|{#2}\Bigr\|^2_{#1}}% norm squared
\newcommand{\iprod}[3][{}]{\langle{#2},{#3}\rangle_{#1}}  % inner product
\newcommand{\bd}  {\partial}                % symbol for boundary of a set
\newcommand{\bigdcup}{\stelldrueber[.45ex]%
   {$\scriptscriptstyle \! \bullet\mspace{1mu}$}{$\bigcup$}}
\newcommand{\restr}[1]{{\restriction}_{#1}} % symbol for map restriction
\newcommand{\map}[3]{{#1}\colon{#2}\longrightarrow{#3}} % maps
\newcommand{\set}[2]{\{ \, #1 \, | \, #2 \, \} } % set

\usepackage{bbm}         % blackboard (see: /usr/share/texmf/mytex/bbm.dvi)
\newcommand{\one}{\mathbbm 1}                    % blackboard 1
\newcommand{\Neu}{{\mathrm N}}              % symbol for Neumann bd cond
\newcommand{\Dir}{{\mathrm D}}              % symbol for Dirichlet bd cond

\newcommand{\laplacian}[1]{\Delta_{{#1}}}   % symbol for Laplacian on mfd
\newcommand{\laplacianD}[1]{\Delta^\Dir_{{#1}}}% symb f Dir-Laplacian
\newcommand{\laplacianT}[1]{\Delta^{\theta}_{{#1}}}% symb f theta-Laplacian
\newcommand{\EW}[2]{\lambda_{#1}({#2})}     % Eigenvalue of Laplacian on #2

\newcommand{\EWD}[2]{\lambda^\Dir_{#1}({#2})}% EV of Dir Laplacian
\newcommand{\EWN}[2]{\lambda^\Neu_{#1}({#2})}% EV of Neu Laplacian
\newcommand{\EWT}[2]{\lambda^{\theta}_{#1}({#2})}% EV of theta Laplacian

\newcommand{\Mnull}{{M_0}}                    % symbol for graph
\newcommand{\Meps}{{M_\eps}}                  % symbol for thickened graph

\newcommand{\Ij}{{I_j}}                       % simple edge
\newcommand{\Ijk}{{I_{jk}}}                   % half simple edge
\newcommand{\pIjk}{{I_{jk}^+}}                % extended interval of edge nbhd
\newcommand{\nI} {{I^0}}                      % added interval
\newcommand{\nIjk}{{I_{jk}^0}}                % eps-indep

\newcommand{\Ueps}{{U_\eps}}                  % with metric
\newcommand{\Uepsj}{{U_{\eps, j}}}            % more precise: with index
\newcommand{\Uepsjk}{{U_{\eps, jk}}}          % half of the edge
\newcommand{\Uj}{{U_j}}                       % only as mfd, with index
\newcommand{\Ujk}{{U_{jk}}}                    % half of the edge only as mfd
\newcommand{\tUeps}{{\tilde U_\eps}}          % with cylindrical metric
\newcommand{\tUepsj}{{\tilde U_{\eps, j}}}    % more precise: with index
\newcommand{\tUepsjk}{{\tilde U_{\eps, jk}}}    % more precise: with index

\newcommand{\pUjk}{{U_{jk}^+}}                % only as mfd, with index

\newcommand{\Veps}{{V_\eps}}                  % with metric
\newcommand{\Vepsk}{{V_{\eps, k}}}            % more precise: with index
\newcommand{\Vk}{{V_k}}                       % only as manifold

\newcommand{\Vnullk}{{V_{0, k}}}         % more precise: with index

\newcommand{\mVeps}{{V_\eps^-}}               % with metric
\newcommand{\mVepsk}{{V_{\eps, k}^-}}         % more precise: with index
\newcommand{\mV}{{V^-}}                       % only as manifold
\newcommand{\mVk}{{V_k^-}}                    % only as manifold

\newcommand{\Aeps}{A_\eps}                    % with metric
\newcommand{\tAeps}{\tilde A_\eps}            % with product metric
\newcommand{\Aepsjk}{A_{\eps, jk}}            % more precise: with index
\newcommand{\Ajk}{A_{jk}}                     % only as manifold

\begin{document}

\title{Convergence of spectra of graph-like thin manifolds}

\author{Pavel Exner}    
\address{Department of Theoretical Physics, NPI, Academy of Sciences,
25068 \v{R}e\v{z} near Prague, and Doppler Institute, Czech
Technical University, B\v{r}ehov\'{a}~7, 11519 Prague, Czechia}
%\email{exner@ujf.cas.cz}
\date{\today}

\author{Olaf Post}   
\address{Institut f\"ur Reine und Angewandte Mathematik,
       Rheinisch-Westf\"alische Technische Hochschule Aachen,
       Templergraben 55,
       52062 Aachen,
       Germany}
%\email{post@iram.rwth-aachen.de}
\date{\today}

%------------------------------------------------------------
% Subject classifications
%------------------------------------------------------------
%Subjclass contains the Classification of the paper following the 1991
%Mathematics Subject Classifications

%\subjclass{35P20, 58G18, 47F05}
%\keywords{Eigenvalues, spectral gap, perturbation of periodic structures}

%------------------------------------------------------------
% Abstract.
%------------------------------------------------------------

\begin{abstract}
  We consider a family of compact manifolds which shrinks with
  respect to an appropriate parameter to a graph. The main result
  is that the spectrum of the Laplace-Beltrami operator converges
  to the spectrum of the (differential) Laplacian on the graph with
  Kirchhoff boundary conditions at the vertices. On the other
  hand, if the shrinking at the vertex parts of the manifold
  is sufficiently slower comparing to that of the edge parts, the
  limiting spectrum corresponds to decoupled edges with Dirichlet
  boundary conditions at the endpoints. At the borderline between
  the two regimes we have a third possibility when the limiting
  spectrum can be described by a nontrivial coupling at the vertices.
\end{abstract}

\maketitle

%------------------------------------------------------------
% The main part
%------------------------------------------------------------

\section{Introduction}

Graph models of quantum systems have a long history. Already half 
a century ago Ruedenberg and Scherr \cite{ruedenberg-scherr:53} 
used this idea, following a suggestion by L.~Pauling, to 
calculate spectra of aromatic carbohydrate molecules; they 
achieved a reasonable accuracy for such a simple model. However, 
a real boom started only from the late eighties when 
semiconductor graph-type structures became small and clean enough 
so that coherent effects in the corresponding quantum transport 
can play the dominating r\^ole. Due to the rapid progress in 
fabrication techniques new systems of these type appear every 
year, making both analysis of graph model properties and their 
physical justification an urgent task. For the sake of brevity we 
avoid giving a review of the field with the list of applications 
and restrict ourselves to quoting the surveys in 
\cite{kostrykin-schrader:99}, \cite{AGHH}, \cite{kuchment:02}, 
\cite{Ku04}. 

Our aim in the present paper is to contribute to the 
understanding of ways in which a graph-type description arises 
from investigation of some really existing systems. To explain 
what we have in mind, recall that the free Hamiltonian of a graph 
model is the (differential) Laplacian on the (metric) graph. To 
define it properly one has to specify the boundary conditions 
which couple the wave functions at the vertices. They have to 
define a self-adjoint operator, however, this requirement itself 
does not specify the conditions uniquely: in a vertex joining $n$ 
graph edges we have $n^2$ free parameters, as observed first in  
\cite{exner-seba:89}.

This non-uniqueness represents the main weakness of graph models.  A natural
idea to mend it is to regard the model in question as a limit case of a more
realistic one with a unique Hamiltonian. An appropriate and natural choice is
a ``thickened graph'' composed of thin tubes which have the same topology as
the original graph and reduce to it in the limit of a vanishing tube radius.
Unfortunately, it is by far not easy to see what happens with spectral and/or
scattering properties in such a limit. After a decade-long effort, the
spectral convergence in the case when the ``thick graph'' is planar with
Neumann boundary conditions has been solved recently by Kuchment--Zeng
\cite{kuchment-zeng:01}, and Rubinstein--Schatzman
\cite{rubinstein-schatzman:01}; Saito~\cite{saito:00} showed the convergence
of the resolvent.\footnote{A related earlier result can be found in the work
  of Colin de Verdi\`ere~\cite{colin:86b} who used the spectral convergence of
  thickened graphs to prove that the first non-zero eigenvalue of a compact
  manifold of dimension greater than $2$ can have arbitrary high (finite)
  multiplicity.}

With the mentioned application to description of quantum wire systems in mind
it is clear that an analogous situation in which the tube boundaries are
Dirichlet is even more important.  Unfortunately, it also very difficult and
despite numerous efforts it remains an open problem.

The main insight of the present paper is that these two cases do not exhaust
all possible ways in which a family of manifolds can approach a graph. One
more choice are manifolds without a boundary of codimension one in $\R^\nu,\:
\nu\ge 3$, which encloses the graph like a system of ``sleeves''\footnote{In
  this context Kuchment and Zeng \cite{kuchment-zeng:01} speak also about
  sleeves. By this notion they mean graph edges thickened into strips. What we
  have in mind here is rather a cylindrical surface with the graph edge as its
  axis.}, with the limit consisting of the sleeve diameter shrinking. It is
not only a mathematical question; we draw the reader's attention to the fact
that such sleeve-shaped tube systems are particularly interesting from the
viewpoint of recent efforts to build circuits based on carbon nanotubes.
Recall that recently discovered techniques --- see, e.g., \cite{andriotis:01,
  papadopoulos:00, terrones:02} --- allow to fabricate branched nanotubes and
thus in principle objects very similar to the mentioned ``sleeved graphs''.

On the mathematical side the main contribution of the paper is the treatment
of the limit problem in a more abstract setting which covers the ``strip
graphs'' of \cite{kuchment-zeng:01, rubinstein-schatzman:01} and their
generalizations to higher dimensions, as well as the ``sleeved graphs''
described above.  This is achieved by using the internal geometry of such a
manifold only, so we need not suppose the latter is embedded in a Euclidean
space. Our results even show that the limit is \emph{independent} of a
particular embedding. Only the abstract graph data count.

Our conclusion will be that such a graph limit can give meaning to some type
of vertex couplings, in particular, those representing a free motion through
the junctions, as well as those which require to extend the graph state space
by extra dimensions corresponding to the vertices. To get the full richness of
the vertex behaviour, with possible relation to the graph geometry, more
general limits will be needed. To characterize the results as well as the
motivation in more details, we need some preliminaries; we will do that in
Sections~\ref{ssec:laplacian.graph} and~\ref{ssec:motivation}.

Finally we give an application on the spectral convergence result in the case
of periodic graphs. In particular, we show the existence of gaps in the
spectrum of certain non-compact periodic graph-like manifolds. For example,
attaching a loop at each vertex gives rise to spectral gaps (cf.\ 
Thm.~\ref{thm:gaps.ex.loop}).

Let us briefly describe the structure of the paper. In
Section~\ref{sec:prelim} we define the Laplacian on a graph and give an
abstract eigenvalue comparison tool (Lemma~\ref{lem:main}). In
Section~\ref{sec:graph.mfd} we define the graph like manifolds associated to a
graph. In Subsection~\ref{ssec:motivation} we motivate the four different
limiting procedures on the vertex neighbourhoods discussed in
Sections~\ref{sec:fast.decay}~--~\ref{sec:alpha.null}. Our main results are
given in Thms.~\ref{thm:ev.conv}, \ref{thm:ev.conv.slow},
\ref{thm:ev.conv.border} and~\ref{thm:ev.conv.null}. In
Section~\ref{sec:edge.nbh} we define the limit procedure of the edge
neighbourhoods which remain the same in all cases.  The last section
(Sec.~\ref{sec:applications}) contains the mentioned applications to periodic
graphs.

%------------------------------------------------------------
\section{Preliminaries}
\label{sec:prelim}
%------------------------------------------------------------

%------------------------------------------------------------
\subsection{Laplacian on a graph}
\label{ssec:laplacian.graph}

Suppose $\Mnull$ is a finite connected graph with vertices $v_k$, $k \in K$
and edges $e_j$, $j \in J$.  Suppose furthermore that $e_j$ has length
$\ell_j>0$, i.e., $e_j \cong \Ij:=[0,\ell_j]$. We clearly can make $\Mnull$
into a metric measure space with measure given by $p_j(x) dx$ on the edge
$e_j$ where $\map {p_j} {\Ij} {(0,\infty)}$ is a smooth density function for
each $j \in J$. We then have
% ------------- %
\begin{gather*}
  \Lsqr \Mnull = \bigoplus_{j \in J} \Lsqr {\Ij, p_j(x)dx}\\
  \normsqr[\Mnull] u = \sum_{j \in J} \normsqr[\Ij] {u_j} =
  \sum_{j \in J} \int_{\Ij} |u_j(x)|^2 p_j(x)dx.
\end{gather*}
% ------------- %
We let $\Sob \Mnull$ be the completion of
% ------------- %
\begin{displaymath}
  \set{u \in \Cont \Mnull} {u_j := u \restr {e_j} \in \Cont[1]{\Ij}}
\end{displaymath}
% ------------- %
where the closure is taken with respect to the norm
% ------------- %
\begin{displaymath}
  \normsqr[1,\Mnull] u :=
  \sum_{j \in J} (\normsqr[\Ij] {u_j} + \normsqr[\Ij]{u_j'}).
\end{displaymath}
% ------------- %
Note that the weakly differentiable functions $\Sob {\Ij}$ on an
interval are continuous, therefore $\Sob \Ij \subset \Cont
\Ij$.

Next we associate with the graph a positive quadratic form,
% ------------- %
\begin{displaymath}
  \normsqr[\Mnull] {u'} := \sum_{j \in J} \normsqr[\Ij]{u_j'}
\end{displaymath}
for all $u \in \Sob \Mnull$. It allows us to define the \emph{(differential)
  Laplacian on the (weighted) graph $\Mnull$} as the unique self-adjoint and
non-negative operator $\laplacian \Mnull$ associated with the closed form $u
\mapsto \normsqr[\Mnull] {u'}$ (see \cite[Chapter~VI]{kato:66},
\cite{reed-simon-1} or \cite{davies:96} for details on quadratic forms). In
other words, the operator and the quadratic form are related by
% ------------- %
\begin{equation}
\label{eq:graphform}
  \normsqr[\Mnull] {u'} = \iprod {\laplacian \Mnull u} u
\end{equation}
% ------------- %
for $u \in \Cont[1] \Mnull$ belonging to the domain of $\laplacian
\Mnull$. On the edge $e_j$, the operator $\laplacian \Mnull$ is
given formally by
% ------------- %
\begin{equation}
\label{eq:formaledge}
  \laplacian \Mnull u =
%   - \frac d {dx_j} \bigl( p_j \frac d {dx_j} u_j \bigr).
   - \frac 1 {p_j(x)}( p_j(x) u_j')'.
\end{equation}
% ------------- %
Note that the domain of $\laplacian \Mnull$ consists of all functions
$u \in \Cont \Mnull$ which are twice weakly differentiable on each
edge. Furthermore, each function $u$ satisfies (weighted)
\emph{Kirchhoff boundary conditions}\footnote{This is the usual
  terminology, not quite a fortunate one. The name suggests that the
  probability current at the vertex obeys the conservation law
  analogous to Kirchhoff's law in an electric circuit. While this
  claim is valid, the current conservation requirement is equivalent
  to selfadjointness and thus also satisfied for the other operators
   mentioned below.} at each vertex $v_k$, i.e.,
% ------------- %
\begin{equation}
  \label{eq:kirchhoff}
  \sum_{\text{$j$, $e_j$ meets $v_k$}} p_j(v_k) u_j'(v_k) = 0
\end{equation}
% ------------- %
for all $k \in K$ where the derivative is taken on each edge in the direction
away from the vertex. In particular, we assume Neumann boundary conditions at
a vertex with only one edge emanating.\footnote{This hypothesis is made for
  convenience only and our result will not change if it is replaced by any
  other boundary condition at the ``loose ends'', in particular, by Dirichlet
  or $\theta$-periodic ones (cf.~Section~\ref{ssec:per.graph}).}  If we assume
that $p$ is continuous on $\Mnull$, we can omit the factors $p_j(v_k)$
in~\eqref{eq:kirchhoff}. Note that different values of $p_j(v_k)$ for $j$ can
correspond in our limiting result to different radii of the thickened edges
which are attached to a vertex neighbourhood (see (\ref{eq:def.radius})
below).

As we have mentioned in the introduction there are other
self-adjoint operators which act according to
(\ref{eq:formaledge}) on the graph edges but satisfy different
boundary conditions at the vertices --- see \cite{exner-seba:89,
kostrykin-schrader:99} for details. The corresponding quadratic
forms differ from (\ref{eq:graphform}) by an extra term. In
general there are many admissible boundary conditions; a graph
vertex joining $n$ edges gives rise to a family with $n^2$ real
parameters. An example is represented by the so-called $\delta$
coupling for which the corresponding domain consists of all
functions $u \in \Cont \Mnull$ which are twice weakly
differentiable on each edge, and (\ref{eq:kirchhoff}) is replaced
by
% ------------- %
\begin{equation}
  \label{eq:delta}
  \sum_{\text{$j$, $e_j$ meets $v_k$}} p_j(v_k) u_j'(v_k) = \kappa
  u(v_k)
\end{equation}
% ------------- %
with a fixed $\kappa\in\mathbb{R}$, where $u(v_k)$ is the common value of all
the $u_j(v_k)$ at the vertex. One can ask naturally whether such graph
Hamiltonians can be obtained from a family of graph-shaped manifolds. In
Section~\ref{sec:borderline} we will discuss a particular case of the limiting
procedure leading to the spectrum which --- although it does \emph{not}
correspond to a graph operator with the generalized boundary condition
described above --- is at least \emph{similar} to that with a $\delta$
coupling. The difference is that in the boundary conditions (\ref{eq:delta})
the coupling constant $\kappa$ is replaced by a quantity dependent on the
spectral parameter, the corresponding operator being defined not on $L^2(M_0)$
but on a slightly enlarged Hilbert space ---
cf.~(\ref{def:lim.border})--(\ref{def:delta.spectral}).

In Section~\ref{sec:slow.decay} we obtain another limit operator due to a
different limiting procedure. This operator is again no graph operator with
boundary conditions as above, but decouples and the graph part corresponds to
a fully decoupled operator with Dirichlet boundary conditions at each vertex.

The spectrum of $\laplacian \Mnull$ is purely discrete. We denote
the corresponding eigenvalues by $\EW k {\laplacian \Mnull} = \EW
k \Mnull$, $k \in \N$, written in the ascending order and repeated
according to multiplicity. With this eigenvalue ordering, we can
employ the \emph{min-max principle} (in the present form it can be
found, e.g., in~\cite{davies:96}): the $k$-th eigenvalue of
$\laplacian \Mnull$ is expressed as
% ------------- %
\begin{equation}
  \label{eq:max.min}
  \EW k \Mnull =
  \inf_{L_k} \sup_{u \in L_k \setminus \{0\} }
      \frac {\normsqr{q_0(u)}}{\normsqr u}
\end{equation}
% ------------- %
where the infimum is taken over all $k$-dimensional subspaces
$L_k$ of $\Sob \Mnull$.

%------------------------------------------------------------
\subsection{Comparison of eigenvalues}
%\label{ssec:eigenvalues}

Let us now formulate a simple consequence of the min-max principle
which will be crucial for the proof of our main results. Suppose that
$\HS$, $\HS'$ are two separable Hilbert spaces with the norms $\norm
\cdot$ and $\norm \cdot '$. We need to compare eigenvalues $\lambda_k$
and $\lambda'_k$ of non-negative operators $Q$ and $Q'$ with purely
discrete spectra defined via quadratic forms $q$ and $q'$ on $\mathcal
D \subset \HS$ and $\mathcal D' \subset \HS$. We set $\normsqr[Q,n] u
:= \normsqr u + \normsqr {Q^{n/2}u}$.

% ------------- %
\begin{lem}
\label{lem:main}
  Suppose that $\map \Phi {\mathcal D}{\mathcal D'}$ is a linear map
  such that there exist constants $n_1, n_2 \ge 0$ and
  $\delta_1, \delta_2 \ge 0$ such that
  % ------------- %
  \begin{align}
    \label{eq:est.norm}
    \normsqr u & \le {\norm{\Phi u}'}^2 + \delta_1 \normsqr[Q,n_1] u\\
    \label{eq:est.quad.form}
    q(u) & \ge \, q'(\Phi u) - \delta_2 \normsqr[Q,n_2] u
  \end{align}
  % ------------- %
  for all $u \in \mathcal D$ and that $\mathcal D \subset \dom
  Q^{\max\{n_1,n_2\}/2}$. Then to each $k$ there is
  a positive function $\eta_k$ given by~\eqref{eq:eta.k}  satisfying
  $\eta_k:=\eta(\lambda_k, \delta_1, \delta_2) \to 0$ as
  $\delta_1, \delta_2 \to 0$, such that
  % ------------- %
  \begin{displaymath}
    \lambda_k \ge \lambda_k' - \eta_k.
  \end{displaymath}
  % ------------- %
\end{lem}
% ------------- %
\begin{proof}
Let $\phi_1, \dots, \phi_k$ be an orthonormal system of
eigenvectors corresponding to the eigenvalues $\lambda_1, \dots,
\lambda_k$. For $u$ in the linear span $E_k$ of $\phi_1, \dots,
\phi_k$, we have
% ------------- %
  \begin{equation}
    \label{eq:est.ev}
    \normsqr[Q,n] u \le (1+\lambda_k^n) \normsqr u
  \end{equation}
% ------------- %
and
% ------------- %
  \begin{multline}
  \label{eq:est.rayleigh}
    \frac{q'(\Phi u)}{{\norm {\Phi u}'}^2} - \frac{q(u)}{\normsqr u}
%     =
%     \frac{q(u)}{\normsqr u} \frac{\normsqr u - {\norm {\Phi u}'}^2}
%     {{\norm {\Phi u}'}^2} +
%                     \frac{q'(\Phi u) - q(u)} {{\norm {\Phi u}'}^2}\\
    \le \left(
              \frac{q(u)}{\normsqr u} \delta_1 \normsqr[Q,n_1] u +
              \delta_2                         \normsqr[Q,n_2] u
        \right)
    \frac 1 {{\norm {\Phi u}'}^2}\\
    \le (\lambda_k(1+\lambda_k^{n_1}) \delta_1 +
    (1+\lambda_k^{n_2})\delta_2)
                    \frac {\normsqr u}{{\norm {\Phi u}'}^2}
  \end{multline}
% ------------- %
where we have used \eqref{eq:est.norm} and
\eqref{eq:est.quad.form} to get the first inequality and
\eqref{eq:est.ev} to get the second one. From relation
\eqref{eq:est.norm} we follow
% ------------- %
  \begin{equation}
    \label{eq:est.norm2}
    (1-(1+\lambda_k^{n_1})\delta_1) \normsqr u \le {\norm {\Phi u}'}^2
  \end{equation}
% ------------- %
and thus we can estimate the r.h.s.\ of \eqref{eq:est.rayleigh} by
% ------------- %
  \begin{equation}
    \label{eq:eta.k}
    \eta_k:= \eta(\lambda_k, \delta_1, \delta_2) :=
    \frac {\lambda_k(1+\lambda_k^{n_1}) \delta_1 +
             (1+\lambda_k^{n_2})\delta_2}
                    {1-(1+\lambda_k^{n_1}) \delta_1}
  \end{equation}
% ------------- %
provided $0 \le \delta_1 < 1/(1+\lambda_k^{n_1})$. From
\eqref{eq:est.norm2} we also conclude that $\norm u = 0$ holds if
$\norm {\Phi u}' = 0$, i.e., that $\Phi(E_k)$ is $k$-dimensional.  From
the min-max principle applied to the quadratic form $q'$ we obtain
% ------------- %
  \begin{displaymath}
    \lambda_k' \le
    \sup_{u \in E_k \setminus \{0\}}\frac{q'(\Phi u)}{{\norm{\Phi u}'}^2}
       \le
    \sup_{u \in E_k \setminus \{0\}}\frac{q(u)}{\normsqr u} + \eta_k =
    \lambda_k + \eta_k
  \end{displaymath}
% ------------- %
which is the desired result.
\end{proof}

%------------------------------------------------------------
\section{Graph-like manifolds}
\label{sec:graph.mfd}
%------------------------------------------------------------

%------------------------------------------------------------
\subsection{Laplacian on a manifold}
%\label{ssec:laplacian.mfd}

Throughout this paper we study manifolds of dimension $d \ge 2$.
For a Riemannian manifold $X$ (compact or not) without boundary we
denote by $\Lsqr X$ the usual $L_2$-space of square integrable
functions on $X$ with respect to the volume measure $dX$ on $X$.
In a chart, the volume measure has the density $(\det G)^{1/2}$
with respect to the Lebesgue measure, where $\det G$ is the
determinant of the metric tensor $G:=(g_{ij})$ in this chart. The
norm of $\Lsqr X$ will be denoted by $\norm[X] \cdot$. For $u \in
\Cci X$, the space of compactly supported smooth functions, we set
% ------------- %
\begin{displaymath}
  \check q_X(u):=\normsqr[X] {d u} = \int_{X} |d u|^2 dX.
\end{displaymath}
% ------------- %
Here the $1$-form $d u$ denotes the exterior derivative of $u$ whose squared
norm in coordinates is given by
% ------------- %
\begin{displaymath}
  |d u |^2= \sum_{i,j} g^{ij} \partial_i u \, \partial_j
\overline u = G^{-1} \nabla u \cdot \nabla \overline u
\end{displaymath}
% ------------- %
where $(g^{ij})$ is the component representation of the inverse
matrix $G^{-1}$.

We denote the closure of the non-negative quadratic form $\check
q_X$ by $q_X$.  Note that the domain $\dom q_X$ of the closed
quadratic form $q_X$ consists of functions in $L_2(X)$ such that
the weak derivative $d u$ is also square integrable, i.e., $q_X(u)
< \infty$.

We define the \emph{Laplacian} $\laplacian X$ (for a manifold
without boundary) as the unique self-adjoint and non-negative
operator associated with the closed quadratic form $q_X$
as in~\eqref{eq:graphform}.

If $X$ is a compact manifold with piecewise smooth boundary $\bd X \ne
\emptyset$ we can define the Laplacian with Neumann boundary condition via the
closure $q_X$ of the quadratic form $\check q_X$ defined on $\Ci X$, the space
of smooth functions with derivatives continuous up to the boundary of $X$.
Note that the usual conditions on the normal derivative occurs only in the
\emph{operator} domain via the Gauss-Green formula. In a similar way other
boundary conditions at $\bd X$ may be introduced. The spectrum of $\laplacian
X$ (with any boundary condition if $\bd X \ne \emptyset$) is purely discrete
as long as $X$ is compact and the boundary conditions are local. We denote the
corresponding eigenvalues by $\EW k {\laplacian X} = \EW k X$, $k \in \N$,
written in increasing order and repeated according to multiplicity.

%------------------------------------------------------------
\subsection{General estimates on manifolds}
%\label{ssec:gen.est}
We will employ (partial) averaging processes on edge and vertex
neighbourhoods which correspond to projection onto the lowest
(transverse) mode.  We start with such a general Poincar\'{e}-type
estimate:
% ------------- %
\begin{lem}
  \label{lem:minmax.2nd.neu}
  Let $X$ be a connected, compact manifold with smooth boundary $\bd
  X$. For $u \in \Sob X$ define the constant function $u_0(x) := \frac
  1 {\vol X} \int_X u\, dX$. Then we have $\normsqr[X]{u_0} \le
  \normsqr[X] u$,
  % ------------- %
  \begin{displaymath}
    \normsqr[X] {u - u_0} \le \frac 1 {\EWN 2 X} \normsqr[X] {du} 
       \qquad \text{and} \qquad
    \normsqr[X] u - \normsqr[X] {u_0} \le
      \frac 1 {\delta \EWN 2 X} + \delta \normsqr[X] u
  \end{displaymath}
  % ------------- %
  for $\delta>0$.
\end{lem}
% ------------- %
\begin{proof}
  The first inequality follows directly from Cauchy-Schwarz. For the
  second one, note that $u - u_0$ is orthogonal to the first
  eigenfunction of the Neumann Laplacian. By the min-max principle we
  obtain
  % ------------- %
  \begin{displaymath}
    \EWN 2 X \normsqr[X] {u - u_0} \le \normsqr[X]{d(u-u_0)} =
    \normsqr[X]{du}. %,
  \end{displaymath}
  % ------------- %
%  because $u_0$ is constant on $X$.
  Since $X$ is connected, we have
  $\EWN 2 X>0$. The last inequality follows from
  \begin{equation}
    \label{ineq:norm}
    | \normsqr u -\normsqr {u_0}| \le
    2 \norm {u - u_0}  \norm u \le
    \frac 1 \delta \normsqr{u - u_0} + \delta \normsqr u
  \end{equation}
  for all $\delta>0$.
\end{proof}
Next, we need the following continuity of the map which restricts a function
on $X$ to the boundary $\bd X$. To this aim we use standard Sobolev embedding
theorems:
% ------------- %
\begin{lem}
  \label{lem:rest.est}
  There exists a constant $c_1>0$ depending only on $X$ and
  the metric $g$ such that
  % ------------- %
  \begin{displaymath}
    \normsqr[\bd X] {u \restr {\bd X}} \le
    c_1(\normsqr[X] u + \normsqr[X] {d u})
  \end{displaymath}
  % ------------- %
  for all $u \in \Sob U$.
\end{lem}
% ------------- %
\begin{proof}
  See e.g. \cite[Ch.~4, Prop.~4.5]{taylor:96}. An alternative proof similar to
  the proof of Lemma~\ref{lem:poincare} exists, and follows easily
  from~\eqref{eq:rest.est} together with a cut-off function.
\end{proof}

%------------------------------------------------------------
\subsection{Definition of the graph-like manifold}
%\label{ssec:def.mfd}

\begin{figure}[h]
  \begin{center}
%------------------------------------------------------------
%    \input{mfd.pstex_t}
\begin{picture}(0,0)%
  \includegraphics{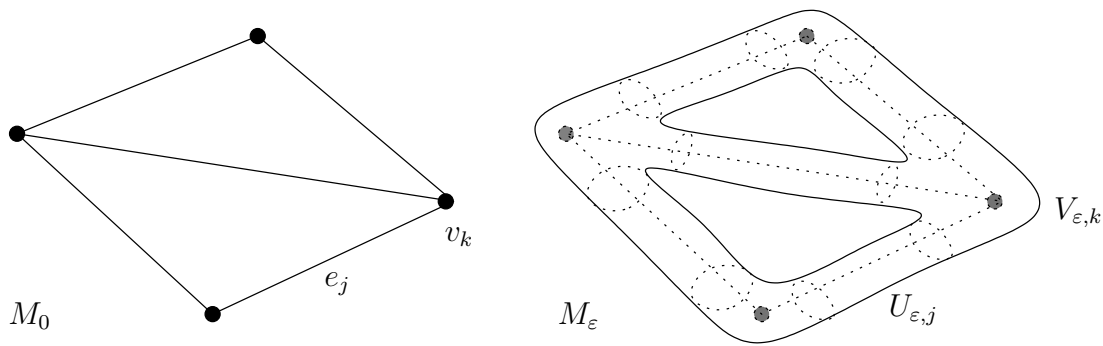}%
\end{picture}%
\setlength{\unitlength}{4144sp}%
\begin{picture}(6263,2015)(308,-1370)
  \put(6571,-601){$\Vepsk$}%
  \put(5581,-1186){$\Uepsj$}%
  \put(3601,-1231){$\Meps$}%
  \put(2926,-736){$v_k$}%
  \put(316,-1231){$\Mnull$}%
  \put(2206,-1006){$e_j$}%
\end{picture}
%------------------------------------------------------------
    \caption{On the left, we have the graph $\Mnull$, on the right, the
      associated graph-like manifold (in this case, $F=\Sphere^1$ and $\Meps$
      is a $2$-dimensional manifold).}
    \label{fig:mfd}
  \end{center}
\end{figure}

For each $0 < \eps \le 1$ we associate with the graph $\Mnull$ a compact and
% ------------- %
\begin{figure}[h]
  \begin{center}
%------------------------------------------------------------
%    \input{edge.vertex.pstex_t}
\begin{picture}(0,0)%
 \includegraphics{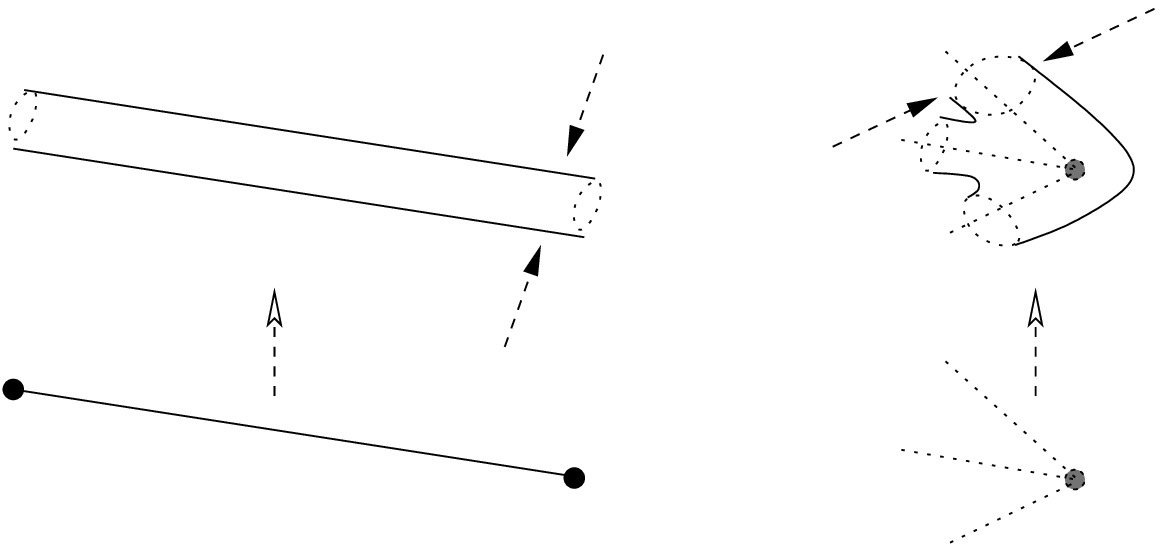}%
\end{picture}%
\setlength{\unitlength}{4144sp}%
\begin{picture}(5282,2500)(374,-1970)
  \put(1591,-586){$\Uepsj$}%
  \put(1561,-1674){$e_j$}%
  \put(5641,-331){$\Vepsk$}%
  \put(5416,-1771){$v_k$}%
  \put(4579,374){$\eps$}%
  \put(3169,-436){$\eps$}%
\end{picture}
%------------------------------------------------------------
    \caption{The associated edge and vertex neighbourhoods with
      $F=\Sphere^1$, i.e., $\Uepsj$ and $\Vepsk$ are $2$-dimensional manifolds
      with boundary.}
    \label{fig:edge.vertex}
  \end{center}
\end{figure}
% ------------- %
connected Riemannian manifold $\Meps$ of dimension $d \ge 2$ equipped with a
metric $g_\eps$ to be specified below. We suppose that $\Meps$ is the union of
compact subsets $\Uepsj$ and $\Vepsk$ such that the interiors of $\Uepsj$ and
$\Vepsk$ are mutually disjoint for all possible combinations of $j \in J$ and
$k \in K$. We think of $\Uepsj$ as the thickened edge $e_j$ and of $\Vepsk$ as
the thickened vertex $v_k$ (see Figures~\ref{fig:edge.vertex}
and~\ref{fig:mfd}).  Note that Figure~\ref{fig:mfd} describes the situation
only roughly, since it assumes that $\Meps$ is embedded in $\R^\nu$. More
correctly, we should think of $\Meps$ as an abstract manifold obtained by
identifying the appropriate boundary parts of $\Uepsj$ and $\Vepsk$ via the
connection rules of the graph $M_0$.
% ------------- %
% ------------- %
This manifold need not to be embedded, but the situation when $\Meps$ is a
submanifold of $\R^\nu$ ($\nu \ge d$) can be viewed also in this abstract
context (see Example~\ref{ex:embedded}).

As a matter of convenience we assume that $\Uepsj$ and $\Vepsk$ are
independent of $\eps$ as manifolds, i.e., only their metric
$g_\eps$ depend on $\eps$.  This can be achieved in the following
way: for the edge regions we assume that $\Uepsj$ is diffeomorphic
to $\Ij \times F$ for all $0 < \eps \le 1$ where $F$ denotes a
compact and connected manifold (with or without a boundary) of
dimension $m:=d-1$.  For the vertex regions we assume that the
manifold $\Vepsk$ is diffeomorphic to an $\eps$-independent
manifold $\Vk$ for $0 < \eps \le 1$. Pulling back the metrics to
the diffeomorphic manifold we may assume that the underlying
differentiable manifold is independent of $\eps$. Therefore,
$\Uepsj = \Uj = \Ij \times F$ and $\Vepsk = \Vk$ with an
$\eps$-depending metric $g_\eps$.

For further purposes, we need a decomposition of 
$e_j \cong \Ij$ into two halves. We reverse the orientation of one
such half so that each half is directed away from its adjacent vertex and
collect all halves $\Ijk$ ending at the vertex $v_k$, i.e., $j \in
J_k$, where
% ------------- %
\begin{equation}
  \label{eq:def.Jk}
  J_k := \set{j \in J}{\text{$e_j$ meets $v_k$}}\footnote{For each loop $e_j$
  at $v_k$, i.e., each edge beginning \emph{and} ending at $v_k$, we need to
  replace the label $j$ by two distinct labels $j_1,j_2$ belonging to $J_k$ in
  order to collect \emph{both} halfs of the edge.}
\end{equation}
% ------------- %
We denote $\Ujk := \Ijk \times F$ (and similar notation with
subscript $\eps$).

For further references, we denote the midpoint of the edge $e_j
\cong \Ij$ by $x_j^*$ and the endpoint of $\Ij$ corresponding to the edge
$v_k$ by $x_{jk}^0$, e.g., $\Ijk = [x_j^*, x_{jk}^0]$.

%------------------------------------------------------------
\subsection{Notation}

In the sequel, we are going to suppress the edge and vertex
subscripts $j$ and $k$ unless a misunderstanding may occur.
Similarly we set, e.g., $U:= U_1$, in other words we omit the
subscript $\eps$ if we only mean the underlying $\eps$-independent
manifold with metric $g_1$, i.e., if we fix $\eps = 1$.

%------------------------------------------------------------
\subsection{Motivation for the different limit operators}
\label{ssec:motivation}
Let us briefly motivate why the limit operator of $\laplacian \Meps$ as $\eps
\to 0$ should depend on the volume decay of the vertex neighbourhoods $\Vepsk$
in comparison with $\vol_{d-1} \bd \Vepsk$ (or $\vol_d \Uepsj$, which is of
the same order when $\eps \to 0$ as we will see in
Section~\ref{sec:edge.nbh}).  For simplicity, we assume that the radius of the
transversal direction on the edge $\Uepsj$ is $\eps$ (i.e., $p_j \equiv 1$).
The assumptions on the edge neighbourhoods will be specified in the next
section. We stress that our aim in this subsection is to present a heuristic
idea, not a proof (for a suitable reasoning cf.\ \cite{ruedenberg-scherr:53}
or \cite{kuchment:02}).

Suppose $\phi=\phi_\eps$ is an eigenfunction of $\laplacian \Meps$ w.r.t the
eigenvalue $\lambda=\lambda_\eps$. By the Gauss-Green formula, we have at the
vertex $\Veps=\Vepsk$
\begin{equation}
  \label{eq:gauss-green}
  \lambda \int_\Veps \phi \, \overline u \, d\Veps =
  \int_\Veps \iprod {d\phi}{du} d\Veps +
  \int_{\bd \Veps} \partial_\mathrm{n} \phi \, \overline u \, d\partial \Veps
\end{equation}
for all $u \in \Sob \Meps$. Assume that $\lambda_\eps \to \lambda_0$ and
$\phi_\eps \to \phi_0=(\phi_{0,j})_j$.

If the vertex volume $\vol_d \Veps$ decays faster than the boundary area
$\vol_{d-1} \bd \Veps$ only the boundary integral over $\bd \Veps$ survives in
the limit $\eps \to 0$ and leads to
\begin{displaymath}
  0=\sum_{j \in J_k} \phi_{0,j}'(v_k)
\end{displaymath}
which is exactly the Kirchhoff boundary condition mentioned above
in~\eqref{eq:kirchhoff}. This \emph{fast decaying} vertex volume case will be
treated in Section~\ref{sec:fast.decay}.

If the vertex volume decays slower than $\vol_{d-1} \bd \Veps$, the integrals
over $\Veps$ are dominant. In this case, $\vol \Vepsk \gg \vol \Uepsj$ and
only slowly varying eigenfunctions on $\Vepsk$ lead to bounded eigenvalues
$\lambda=\lambda_\eps$. Since $\vol \Vepsk \gg \vol \Uepsj$, normalized
eigenfunctions are nearly vanishing on $\Vepsk$ viewed from the scale on
$\Uepsj$.  This roughly explains, why we end up with a decoupled operator with
Dirichlet boundary conditions on $\Mnull$ plus extra zero eigenmodes at the
vertices (the zero eigenmodes also survive the limit $\eps \to 0$). This
\emph{slowly decaying} vertex volume case will be discussed in
Section~\ref{sec:slow.decay}.

In the borderline case when $\vol_d \Veps \approx \vol_{d-1} \bd \Veps$, we
also expect the eigenfunctions to vary slowly on $\Vepsk$ (since $\vol_d
\Vepsk \to 0$), so the integral over $\iprod {d\phi}{du}$ should tend to $0$,
and in the limit
\begin{displaymath}
  \lambda_0 \phi_0(v_k) = \sum_{j \in J_k} \phi_{0,j}'(v_k). 
\end{displaymath}
This \emph{borderline case} will be treated in Section~\ref{sec:borderline}.

If $\vol \Vepsk$ does not tend to $0$, i.e., when $\Vepsk$ tends to a compact
$d$-dimensional manifold $\Vnullk$ without boundary (and \emph{not} to a point
as in the cases above), we still expect a decoupled operator with Dirichlet
boundary conditions on the edges by the same arguments as in the slowly
decaying case. In addition, not only the lowest eigenmode of $\Vepsk$ but all
eigenmodes survive, i.e., the limit operator should consist of the direct sum
of all Dirichlet Laplacians on the edges plus the Laplacians on $\Vnullk$, $k
\in K$.  This \emph{non-decaying} vertex volume case will be treated in
Section~\ref{sec:alpha.null}.

It requires an extra effort to prove rigorously the conclusions of the above
reasoning; recall that we have assumed e.g.\ that $\lambda_\eps \to \lambda_0$
(which we want to show in this paper) and
$\norm[\infty]{\phi_\eps},\norm[\infty]{d \phi_\eps} \le c$ which is in
general not true for normalized ($L_2$-)eigenfunctions since $\vol \Meps \to
0$ as $\eps \to 0$.

%------------------------------------------------------------
\section{Edge neighbourhoods}
\label{sec:edge.nbh}
%------------------------------------------------------------

%------------------------------------------------------------
\subsection{Definition of the thickened edges}
%\label{ssec:def.edge}
Suppose that $U=I \times F$ with metric $g_\eps$, where $I$ corresponds to
some (part of an) edge and $F$ denotes (as before) a compact and connected
Riemannian manifold of dimension $m=d-1$ with metric $h$ with or without
boundary (we always assume that the corresponding Laplacian on $\Meps$
satisfies Neumann boundary conditions on the boundary part coming from $\bd
F$).  For simplicity we assume that $\vol F = 1$. We define another metric
$\tilde g_\eps$ on $\Ueps$ by
% ------------- %
\begin{equation}
  \label{eq:def.met.edge}
  \tilde g_\eps := dx^2 + \eps^2 r_j^2(x) \, h(y), \qquad 
  (x,y) \in \Uj = \Ij \times F
\end{equation}
% ------------- %
where
\begin{equation}
  \label{eq:def.radius}
  r_j(x):=(p_j(x))^{1/m}
\end{equation}
defines a smooth function (specifying the radius of the fibre $\{x\} \times F$
at the point $x$), where $p_j$ is the density function on the edge $e_j$
introduced in Section~\ref{sec:prelim}.

We denote by $G_\eps$ and $\tilde G_\eps$ the $d \times d$-matrices associated
to the metrics $g_\eps$ and $\tilde g_\eps$ with respect to the coordinates
$(x,y)$ (here $y$ stands for suitable coordinates on $F$) and assume that the
two metrics coincide up to an error term as $\eps\to 0$, more specifically
% ------------- %
\begin{equation}
  \label{eq:asym.met.edge}
  G_\eps = \tilde G_\eps +
    \begin{pmatrix}
      o(1) & o(\eps) \\ o(\eps) &
      o(\eps^2)
    \end{pmatrix} =
    \begin{pmatrix}
      1 + o(1) & o(\eps) \\ o(\eps) &
      \eps^2 r_j \,  H + o(\eps^2)
    \end{pmatrix},
\end{equation}
% ------------- %
i.e.,
% ------------- %
\begin{displaymath}
  g_{\eps,xx} = 1 + o(1), \quad
  g_{\eps,xy_\alpha} = o(\eps), \quad
  g_{\eps,y_\alpha y_\beta} = 
      \eps^2 r_j^2(x) \, h_{\alpha\beta}(y) + o(\eps^2).
\end{displaymath}
uniformly on $U$. To summarize, we assume that the metric $g_\eps$ is equal to
the product metric $\tilde g_\eps$ up to error terms.

This is a central assumption in our construction which describes how in fact
the family of manifolds shrinks to the graph $M_0$. One of the reasons why we
introduce a pair of metrics will become clear in the following two examples.
While our construction uses intrinsic metric properties of the manifolds only,
we want it to be applicable to manifolds embedded into some Euclidean space
$\R^\nu$. It will be one of our aims to show that within the prescribed error
margin such a ``practically important'' metric yields the same result as the
product metric which is easier to handle.

In particular, our results show that the convergence is \emph{independent} of
the chosen embedding.

%------------------------------------------------------------
\begin{exmp}
  \label{ex:embedded}
  \textbf{Embedded graphs.} Note that it is impossible to embed our graph
  neighbourhood $\Meps$ if the cylindrical sleeves have the \emph{same} length
  as the underlying graph edges, but it can be achieved with the length
  \emph{shortened} by a factor of order $o(1)$. In this sense, we recover the
  situation treated in \cite{kuchment-zeng:01}, i.e., $\Mnull$ embedded in
  $\R^2$ and $F=[-1,1]$ and $\Meps$ being a suitable $\eps$-neighbourhood of
  $\Mnull$.
  
  In the same way, we can treat the graph $M_0$ embedded in $\R^3$ with
  $\Meps$ being the surface of some pipeline network (i.e.\ $F=\Sphere^1$).
\end{exmp}

\begin{exmp}
  \label{ex:curved}
  \textbf{Curved edges and variable transversal radius.} Suppose $\Ueps$ is
  the $\eps$-neighbourhood of a smooth curve $\map{\vec \gamma=\vec
    \gamma_j}{\Ij}{\R^d}$ parameterized by arc-length. If, e.g., $\nu=d=2$ and
  $F=[-1,1]$ a chart is given by
  % ------------- %
  \begin{displaymath}
    \map \Psi {\Ij \times [-1,1]} \Uepsj, \qquad
    (x,y) \mapsto \vec \gamma(x) + \eps r_j(x) y \, \vec n(x),
  \end{displaymath}
  % ------------- %
  i.e., we thicken the curve $\vec \gamma$ in its normal direction $\vec n(x)$
  at the point $\vec \gamma(x)$ by the factor $\eps r(x) = \eps r_j(x)$. The
  corresponding metric in $(x,y)$-coordinates is given by
    % ------------- %
    \begin{displaymath}
      G_\eps =
      \begin{pmatrix}
        (1+ \eps \kappa y r)^2 + \eps^2 y^2\dot r^2 &
        \eps^2 r \dot r y\\ \eps^2 r \dot r y &  \eps^2 r^2
      \end{pmatrix} =
      \begin{pmatrix}
        1+O(\eps) & O(\eps^2) \\ O(\eps^2) &
        \eps^2 r^2
      \end{pmatrix}
    \end{displaymath}
    % ------------- %
    where $\kappa:=\dot \gamma_1 \ddot \gamma_2 - \dot \gamma_2 \ddot
    \gamma_2$ is the curvature of the generating curve $\vec \gamma$.
    Therefore, the error term $o(1)$ comes from the curvature of the embedded
    curve $\vec \gamma$ whereas the off-diagonal error terms come from the
    variable radius of the transversal direction (note that $\dot r=0$ if
    $r(x)$ is constant). Curvature induced effects in the thin tube limit are
    well understood --- see, e.g., \cite{duclos-exner:95}.
\end{exmp}
% ------------- % end of the examples

%------------------------------------------------------------
\subsection{Estimates on the thickened edges}
%\label{ssec:est.edges}

Following the philosophy explained in the previous subsection, we start with
pointwise estimates where we compare the product metric $\tilde g_\eps$ with
the original metric $g_\eps$. Note that the
assumption~\eqref{eq:asym.met.edge}, while fully sufficient for our purposes,
is optimal in a sense, i.e., that the following lemma ceases to be valid if we
weaken its hypothesis even slightly.
% ------------- %
\begin{lem}
\label{lem:metric}
  Suppose that $g_\eps$, $\tilde g_\eps$ are given as
  in~\eqref{eq:def.met.edge} and~\eqref{eq:asym.met.edge}, then
  % ------------- %
  \begin{align}
  \label{eq:met.vol}
    (\det G_\eps)^\frac12 &= (1 + o(1)) (\det \tilde G_\eps)^\frac12 \\
  \label{eq:met.1st.comp}
    g_\eps^{xx} & \!:= (G_\eps^{-1})_{xx} = 1 + o(1) \\
  \label{eq:met.1st.der}
    |d_x u|^2 &\le (1+ o(1)) |d u|_{g_\eps}^2\\
  \label{eq:met.2nd.der}
    |d_F u|_h^2 &\le o(\eps) |d u|_{g_\eps}^2
  \end{align}
  % ------------- %
  where $d_x$ and \ $d_F$ are the (exterior) derivative with respect to
  $x \in I$ and \ $y \in F$, respectively. All the estimates are
  uniform in $(x,y)$ as $\eps \to 0$.
\end{lem}
\begin{proof}
  The first equation follows from
  % ------------- %
  \begin{displaymath}
    \begin{split}
      \det(G_\eps \tilde G_\eps^{-1})
      &= \det
      \begin{pmatrix}
        1 + o(1)  & o(\eps)\\
        o(\eps)   & \eps^2 H + o(\eps^2)
      \end{pmatrix}
      \begin{pmatrix}
        1 & 0\\
        0 & \eps^{-2} H^{-1}
      \end{pmatrix}\\
      &= \det
      \begin{pmatrix}
        1 + o(1)  & o(\eps^{-1})\\
        o(\eps)   & \one + o(1)
      \end{pmatrix}
      = 1 + o(1).
    \end{split}
  \end{displaymath}
  % ------------- %
  For the second one, we consider the upper left component of
  % ------------- %
  \begin{multline*}
%    \begin{split}
      G_\eps^{-1} - \tilde G_\eps^{-1} 
%&
=
      -\tilde G_\eps^{-1} (G_\eps - \tilde G_\eps) \tilde G_\eps^{-1} +
       o(G_\eps - \tilde G_\eps) \\
%&
=
      \begin{pmatrix}
        1 & 0 \\ 0 & O(\eps^{-2})
      \end{pmatrix}
      \begin{pmatrix}
        o(1) & o(\eps) \\ o(\eps) & o(\eps^2)
      \end{pmatrix}
      \begin{pmatrix}
        1 & 0 \\ 0 & O(\eps^{-2})
      \end{pmatrix}
        + o(1) 
%\\ &
     =
     \begin{pmatrix}
       o(1) & o(\eps^{-1}) \\ o(\eps^{-1}) & o(\eps^{-2})
     \end{pmatrix}.
%    \end{split}
  \end{multline*}
  % ------------- %
  Inequality~\eqref{eq:met.1st.der} is equivalent to
  % ------------- %
  \begin{displaymath}
    \begin{pmatrix}
      1 & 0\\ 0 & 0
    \end{pmatrix}
    \le (1+ o(1)) G_\eps^{-1}
  \end{displaymath}
  % ------------- %
  in the sense of quadratic forms. This will be true if we show that
  % ------------- %
  \begin{displaymath}
    \begin{pmatrix}
      1 & 0 \\  0 & \delta \one
    \end{pmatrix}
    \le (1 + o(1)) G_\eps^{-1}
  \end{displaymath}
  % ------------- %
  for some $\delta>0$, where $\one$ is the $m\times m$ unit
  matrix, which in turn means
  % ------------- %
  \begin{displaymath}
    (1+ o(1))
    \begin{pmatrix}
      1 & 0 \\ 0 & \delta^{-1} \one
    \end{pmatrix}
     \ge G_\eps.
  \end{displaymath}
  % ------------- %
  However,
  % ------------- %
  \begin{displaymath}
    G_\eps = \tilde G_\eps +
    \begin{pmatrix}
      o(1) & o(\eps) \\ o(\eps) & o(\eps^2)
    \end{pmatrix}
    =
    \begin{pmatrix}
      1 + o(1) & 0 \\ 0 & O(\eps^2)
    \end{pmatrix}
    +
    \begin{pmatrix}
      0 & o(\eps) \\ o(\eps) & 0
    \end{pmatrix}
  \end{displaymath}
  % ------------- %
  and the eigenvalues of the last matrix are of order
  $o(\eps)$, so
  % ------------- %
  \begin{displaymath}
    G_\eps \le
    \begin{pmatrix}
      1 + o(1) & 0 \\ 0 & O(\eps^2)
    \end{pmatrix}
    + o(\eps) \one =
    \begin{pmatrix}
      1 + o(1) & 0 \\ 0 & o(\eps)
    \end{pmatrix}
    \le
    (1 + o(1))
    \begin{pmatrix}
      1 & 0 \\ 0 & c \one
    \end{pmatrix}
  \end{displaymath}
  % ------------- %
  for some constant $c>0$, and therefore it is sufficient to
  choose $\delta<c^{-1}$.
  The proof of inequality~\eqref{eq:met.2nd.der} is similar.
\end{proof}

%------------------------------------------------------------
\subsection{Notation}
The tilde in a symbol refers always to the product metric $\tilde g_\eps$. We
denote, e.g., by $\tUeps$ the manifold $\Ueps$ with metric $\tilde g_\eps$ and
(abusing the notation a little bit) employ the symbol $\Ueps$ for the manifold
$\Ueps$ with the metric $g_\eps$.

As a motivation for the above choice of the metrics, let us
calculate the norm of $u \in \Lsqr {\tUeps}$ for a function $u$
which is independent of the second argument $y \in F$, i.e.
$u(x,y)=u(x)$. This yields
% ------------- %
\begin{multline}
  \label{eq:ind.of.2nd}
%  \begin{split}
    \normsqr[\tUeps] u 
%& 
=\int_{\Ueps}|u|^2 d\tUeps
%\\
%    &
=\int_I \int_F |u(x)|^2 (\det \tilde G_\eps)^\frac12(x,y) \,dx \,dy \\
%    &
=\eps^m \int_I |u(x)|^2 r^m(x) dx \int_F (\det H)^\frac12(y) \,dy
%\\
%    &
=\eps^m \normsqr[I] u \vol(F) =\eps^m \normsqr[I] u.
%  \end{split}
\end{multline}

%------------------------------------------------------------
\subsection{Transversal averaging}

We will employ averaging processes on edge neighbourhoods $\Ueps=\Uepsj$
which correspond to projection onto the lowest transverse mode:
% ------------- %
\begin{equation}
   \label{def:edge.av}
  N u(x) = N_j u(x) := \int_F u(x,\cdot) \, dF
\end{equation}
% ------------- %
Note that $N u(x)$ is well defined for $u \in \Sob \Ueps$, and
moreover,
% ------------- %
\begin{align}
  \label{eq:edge.av.cont}
  \eps^m \normsqr[I] {N u}  = \normsqr[\tUeps] {N u} &\le
  \normsqr[\tUeps] u = (1+o(1)) \normsqr[\Ueps] u
\end{align}
% ------------- %
in view of eqs.~\eqref{eq:ind.of.2nd},~\eqref{eq:met.vol}, and
the Cauchy-Schwarz inequality.

In the following two lemmas we compare a function $u$ and its
derivative $du$ with the normal averages $Nu$ and \ $d_x Nu$,
respectively. Note that for the next lemma, \eqref{eq:edge.av.cont} is
not enough; we also need the reverse inequality:
% ------------- %
\begin{lem}
  \label{lem:diff.norm.av}
  For any $u \in \Sob \Ueps$ we have
  % ------------- %
  \begin{displaymath}
    \normsqr[\Ueps] u - \normsqr[I] {\eps^{m/2} N u} \le
    o(\eps^{1/2}) \left(\normsqr[\Ueps] u + \normsqr[\Ueps] {du}\right).
  \end{displaymath}
  % ------------- %
\end{lem}
% ------------- %
\begin{proof}
  Applying Lemma~\ref{lem:minmax.2nd.neu} with $X=F$ we get
  % ------------- %
  \begin{equation}
    \label{ineq:fibre}
    \normsqr[F] {u(x,\cdot)} -|N u(x)|^2 \le
    \frac 1 {\delta \EWN 2 F}  \normsqr[F]{d_Fu(x, \cdot)} + 
        \delta \normsqr[F]{u(x, \cdot)}.
  \end{equation}
  % ------------- %
  Next we integrate over $\Ij$ and obtain
  % ------------- %
  \begin{displaymath}
    \normsqr[\tUeps] u - \eps^m \normsqr[\Ij] {N u} \le
    \frac {o(\eps)} {\delta \EWN 2 F}
        \int_{\tUepsj} |d u|^2_{g_\eps} d\tUepsj + \delta \normsqr[\tUepsj] u
  \end{displaymath}
  % ------------- %
  using estimate~\eqref{eq:met.2nd.der}. We put $\delta := \sqrt{o(\eps)}$ and
  apply~\eqref{eq:met.vol} to obtain the result for the manifold $\Uepsj$.
\end{proof}

% ------------- %
\begin{lem}
  \label{lem:diff.quad.av}
  For any $u \in \Sob \Ueps$ we have
  % ------------- %
  \begin{displaymath}
    \normsqr[I] {\eps^{m/2} (Nu)'} - \normsqr[\Ueps] {du} \le
    o(1) \normsqr[\Ueps] {du}.
  \end{displaymath}
  % ------------- %
\end{lem}
% ------------- %
\begin{proof}
% ------------- %
  \begin{multline*}
    \normsqr[I] {\eps^{m/2} (Nu)'} =
    \normsqr[I] {\eps^{m/2} N (d_x u)} \le
    (1+o(1))\normsqr[\Ueps] {d_x u} \\ =
    (1+o(1))\int_{\Ueps} |d_x u|^2 \, d \Ueps \le
    (1+o(1))\int_{\Ueps} |d u|_{g_\eps}^2 \, d \Ueps =
    (1+o(1))\normsqr[\Ueps] {du}
  \end{multline*}
% ------------- %
  holds in view of estimate~\eqref{eq:edge.av.cont} and~\eqref{eq:met.1st.der}.
\end{proof}
% ------------- %

Next we need a pointwise estimate on the behaviour of $Nu$ at the
boundary.
% ------------- %
\begin{lem}
  \label{lem:av.edge}
  We have
  % ------------- %
  \begin{displaymath}
    |N u (x^0)|^2 \le
    O(\eps^{-m}) (\normsqr[\Ueps] u + \normsqr[\Ueps] {du})
  \end{displaymath}
  % ------------- %
  for all $u \in \Sob \Ueps$, where $x^0 \in \bd I$.
\end{lem}
% ------------- %
\begin{proof}
  One estimates
  % ------------- %
  \begin{displaymath}
    |N u (x^0)|^2 \le
    \int_F |u(x^0, y)|^2 \, dF(y) \le
    c_1(\normsqr[U] u + \normsqr[U] {d_x u}) \le
    O(\eps^{-m}) (\normsqr[\Ueps] u + \normsqr[\Ueps] {du})
  \end{displaymath}
  % ------------- %
  by Lemma~\ref{lem:rest.est} with $X=U$,~\eqref{eq:met.1st.der}
  and~\eqref{eq:met.vol}.
\end{proof}

%------------------------------------------------------------
\section{Fast decaying vertex volume}
\label{sec:fast.decay}
%------------------------------------------------------------
\subsection{Definition of the thickened vertices}

Remember that $\Vepsk = \Vk$ as manifold, whereas $g_\eps$ denotes the
$\eps$-depending metric on $\Vepsk$. Let $g:=g_1$, then we assume that
% ------------- %
\begin{equation}
  \label{eq:met.vertex}
  c_- \eps^2 g \le g_\eps \le c_+ \eps^{2\alpha} g
\end{equation}
% ------------- %
in the sense that there are constants $c_-, c_+>0$ such that
% ------------- %
\begin{displaymath}
  c_- \eps^2 g(x)(v,v) \le g_\eps(x)(v,v) \le c_+ \eps^{2\alpha} g(x)(v,v)
\end{displaymath}
% ------------- %
for all $v \in T_x \Vk$ and all $x \in \Vk$. The number $\alpha$
in the exponent is assumed to satisfy the inequalities
% ------------- %
 \begin{equation}
  \label{eq:est.alpha.fast}
  \frac {d-1} d < \alpha \le 1\,;
\end{equation}
% ------------- %
notice that $\alpha \le 1$ is needed for \eqref{eq:met.vertex} to make sense
with $0<\eps\le 1$. Thus the edge and vertex parts of the manifold need not
shrink at the same rate but the vertex shrinking should not be too slow than
that of the edges. This hypothesis expressed by (\ref{eq:met.vertex}) plays a
central r\^ole here; other shrinking regimes will be discussed in the
following sections.

Note that the manifold $\Vepsk$ shrinks at most as $\eps$ (in each direction)
by the lower bound in~\eqref{eq:met.vertex}. This ensures that a global
\emph{smooth} metric $g_\eps$ exists on $\Meps$ with the requirements on
$\Uepsj$ and $\Vepsk$. Therefore, we do not need an intermediate part (called
\emph{bottle neck}) between the edge and vertex neighbourhoods interpolating
between the different scalings as in Sections~\ref{sec:slow.decay}
and~\ref{sec:borderline}. % (see also Remark~\ref{rem:alpha.bigger.1}).

We easily obtain the following global estimate from~\eqref{eq:met.vertex}:

% ------------- %
\begin{lem}
  \label{lem:est.norm.quad}
  There are $c_1^\pm, c_2^\pm > 0$ such that
  % ------------- %
  \begin{align}
    \label{eq:est.vert.norm}
    c_1^- \eps^d \normsqr[V] u & \le
    \normsqr[\Veps] u \le
    c_1^+ \eps^{\alpha d} \normsqr[V] u \\
    \label{eq:est.vert.quad}
    c_2^- \eps^{d-2\alpha} \normsqr[V] {du} & \le
    \normsqr[\Veps] {du} \le
    c_2^+ \eps^{\alpha d-2} \normsqr[V] {du}
  \end{align}
  % ------------- %
  for all $u \in \Sob \Veps = \Sob V$.
\end{lem}

%------------------------------------------------------------
\subsection{Convergence of the spectra}
The limit operator will concentrate only on the edge part in this case,
therefore we define
% ------------- %
\begin{equation}
  \label{eq:lim.slow}
  \HS_0 := \Lsqr \Mnull, \qquad
  \mathcal D_0 := \Sob \Mnull, \qquad
  q_0(u):= \normsqr[\Mnull] {u'} = \sum_j \normsqr[\Ij] {u_j'},
\end{equation}
% ------------- %
i.e., the limit operator $Q_0$ is $\laplacian \Mnull$ (see
Def.~\eqref{eq:formaledge}). If the transversal manifold $F$ has boundary, we
assume that $\laplacian \Meps$ satisfies Neumann boundary conditions. With the
above preliminaries we can finally formulate the main result of this section:
% ------------- %
\begin{thm}
\label{thm:ev.conv}
  Under the stated assumptions $\EW k \Meps \to \EW k \Mnull$
  as $\eps\to 0$.
\end{thm}
% ------------- %
\noindent Recall that the eigenvalues $\EW k \Meps$ are by
assumption ordered in the ascending order, multiplicity taken into account, so
the label of a particular eigenvalue curve may change as $\eps$ moves. The
spectrum of the Laplacian on $\Meps$ is in general richer than that of the
graph and a part of the eigenvalues escapes to $+\infty$ as $\eps\to 0$; the
proof presented below shows that this happens, roughly speaking, for all
states with the transverse part of the eigenfunction orthogonal to the ground
state.

Our aim is to find a two sided estimate on each eigenvalue $\EW k
\Meps$ by means of $\EW k \Mnull$ with an error which is $o(1)$
w.r.t.\ the parameter $\eps$.

%------------------------------------------------------------
\subsection{An upper bound}
%\label{ssec:up.est}

The mentioned upper eigenvalue estimate now reads as follows:
% ------------- %
\begin{thm}
\label{thm:ev.above}
  $\EW k \Meps \le \EW k \Mnull + o(1)$ holds as $\eps\to 0$.
\end{thm}
% ------------- %
\noindent To prove it, we define the transition operator by
% ------------- %
\begin{equation}
  \label{eq:trans.op.above}
  \Phi_\eps u (z):=
  \begin{cases}
    \eps^{-m/2} u(v_k) & \text{if $z \in \Vk$},\\
    \eps^{-m/2} u_j(x) & \text{if $z=(x,y) \in \Uj$}
  \end{cases}
\end{equation}
% ------------- %
for any $u \in \Sob \Mnull$. Theorem~\ref{thm:ev.above} is then
implied by Lemma~\ref{lem:main} in combination with the following
result.

% ------------- %
\begin{lem}
\label{lem:above} We have $\Phi_\eps u \in \Sob \Meps$, i.e.,
$\Phi_\eps$ maps the quadratic form domain of the Laplacian on the
graph into the quadratic form domain of the Laplacian on the
manifold. Furthermore, for $u \in \Sob \Mnull$ we have
% ------------- %
  \begin{align}
    \normsqr[\Mnull] u - \normsqr[\Meps] {\Phi_\eps u} &\le
    o(1) \, \normsqr[\Mnull] u\\
    \normsqr[\Meps] {d \, \Phi_\eps u} - q_0(u) & =
    o(1) \, q_0(u).
  \end{align}
% ------------- %
\end{lem}
% ------------- %
\begin{proof}
  The first assertion is true since $\Phi_\eps u$ is constant on each
  thickened vertex $\Vepsk$ and continuous on $\bd \Vepsk$. Clearly,
  $\Phi_\eps u$ is weakly differentiable on each thickened edge $\Uepsj$.
  Moreover, we have
  % ------------- %
  \begin{multline*}
    \normsqr[\Mnull] u - \normsqr[\Meps] {\Phi_\eps u} \le
    \sum_{j \in J} (\normsqr[\Ij] u - \normsqr[\Uepsj] {\Phi_\eps u}) \\ =
    \sum_{j \in J} (\normsqr[\Ij] u - (1+o(1)) \normsqr[\tUepsj] {\Phi_\eps u})=
    o(1) \sum_{j \in J} \normsqr[\Ij] u = o(1)
    \normsqr[\Mnull] u
  \end{multline*}
  % ------------- %
  where we have neglected the contribution to the norm of $\Phi_\eps u$
  from the vertex parts of $\Meps$ and employed eqs.~\eqref{eq:met.vol}
  and~\eqref{eq:ind.of.2nd}. The second relation follows from
  % ------------- %
  \begin{multline*}
    \normsqr[\Meps] {d \, \Phi_\eps u} - q_0(u) =
    \sum_{j \in J} ((1+o(1))\normsqr[\tUepsj]{g_\eps^{xx} d_x \Phi_\eps u} -
             \normsqr[\Ij]{u'})\\ =
    \sum_{j \in J} ((1+o(1))\normsqr[\Ij] {u'} - \normsqr[\Ij]{u'}) =
    o(1) \, q_0(u)
  \end{multline*}
  % ------------- %
  in the same way as above and with \eqref{eq:met.1st.comp}; recall that
  $\Phi_\eps u$ is constant on $\Vepsk$ and independent of $y \in F$ on
  $\Uepsj$.
\end{proof}
% ------------- %

%------------------------------------------------------------
\subsection{A lower bound}
%\label{ssec:low.est}

The reverse estimate is more difficult. Here, we will also employ
averaging processes on the vertex neighbourhoods $\Vepsk$ which
correspond to projection onto the lowest (constant) mode:
% ------------- %
\begin{equation}
  \label{def:vert.av}
  C u = C_k u := \frac 1 {\vol {\Vk}} \int_{\Vk} u \, d \Vk.
\end{equation}
% ------------- %
Recall that $V=\Vk$ denotes the manifold $\Vk$ with the metric $g=g_1$
(see Remark~\ref{rem:no.eps} for the reason why we use $\Vk$ instead
of $\Vepsk$).

% ------------- %
\begin{lem}
  \label{lem:diff.av}
  The inequality
  % ------------- %
  \begin{displaymath}
    |C_k u - N_j u(x^0)|^2 \le O(\eps^{2\alpha-d}) \normsqr[\Vepsk] {du}
  \end{displaymath}
  % ------------- %
  holds for all $u \in \Sob \Vepsk$ where the point $x^0=x^0_{jk} \in \bd \Ij$
  corresponds to the vertex $v_k$.
\end{lem}
% ------------- %
\begin{proof}
% ------------- %
  \begin{multline*}
    |C_k u - N_j u(x^0)|^2 \le
    \int_F |C_k u - u(x^0,y)|^2 \, dF(y)  \le
    c_1 \left(\normsqr[\Vk] {C_k u - u} + \normsqr[\Vk] {du} \right) \\ \le
    c_1 \left(\frac 1 {\EWN 2 {\Vk}}+1 \right) \normsqr[\Vk] {du}  \le
    O(\eps^{2\alpha-d}) \normsqr[\Vepsk] {du}
  \end{multline*}
  % ------------- %
  holds by Lemma~\ref{lem:rest.est} and Lemma~\ref{lem:minmax.2nd.neu}
  with $X=\Vk$ and metric $g=g_1$, and Lemma~\ref{lem:est.norm.quad}.
\end{proof}

% ------------- %
\begin{lem}
  \label{lem:diff.vol.av}
  We have
  % ------------- %
  \begin{displaymath}
    \normsqr[\Veps] {u - C u} \le O(\eps^\beta) \normsqr[\Veps] {du}
  \end{displaymath}
  % ------------- %
  for all $u \in \Sob \Veps$, where $\beta :=(2+d)\alpha - d$.
\end{lem}
% ------------- %
\begin{proof}
  Using again Lemmas~\ref{lem:minmax.2nd.neu} and~\ref{lem:est.norm.quad} we
  infer
  % ------------- %
  \begin{displaymath}
    \normsqr[\Veps] {u - C u} \le
    c_1^+ \eps^{\alpha d} \normsqr[V] {u - C u} \le
    c_1^+ \eps^{\alpha d} \, \frac 1 {\EWN 2 V} \normsqr[V] {du} \le
   O(\eps^{\alpha d - d + 2\alpha}) \normsqr[\Veps] {du}.
   % ------------- %
  \end{displaymath}
% ------------- %
\noindent Notice that $\beta>0$ is equivalent to $\alpha >
d/(d+2)$ and the last inequality is satisfied due
to~\eqref{eq:est.alpha.fast} and the fact that $d \ge 2$ holds by
assumption.
\end{proof}

% ------------- %
\begin{rem}
  \label{rem:no.eps}
  For Lemma~\ref{lem:diff.vol.av}, the ``natural'' averaging $C_\eps u :=
  \int_\Veps u \,d\Veps$ would yield the same result whereas
  Lemma~\ref{lem:diff.av} leads to the estimate $O(\eps^{\beta-d})$ which is
  worse since $2\alpha>\beta$.
\end{rem}
% ------------- %

We conclude that in the fast decaying case the edge neighbourhoods lead to no
spectral contribution in the limit $\eps \to 0$:
% ------------- %
\begin{cor}
  \label{cor:vertex.small}
  The inequality
  % ------------- %
  \begin{displaymath}
    \normsqr[\Veps] u \le O(\eps^{\alpha d - m})(\normsqr[\Ueps \cup \Veps] u
    + \normsqr[\Ueps \cup \Veps]{du})
  \end{displaymath}
  % ------------- %
  holds true for all $u \in \Sob {\Ueps \cup \Veps}$.
\end{cor}
% ------------- %
\begin{proof}
We start from the telescopic estimate
% ------------- %
  \begin{multline*}
    \norm[\Veps] u \le
    \norm[\Veps] {u - C u} +
      \norm[\Veps] {C u - N u(x^0)} +
      \norm[\Veps] {N u(x^0)} \\ \le
    O(\eps^{\beta/2}) \norm[\Veps]{du} +(\vol \Veps)^{1/2} \left(
    O(\eps^{(2\alpha-d)/2}) \normsqr[\Veps]{du} +  O(\eps^{-m} )
    (\normsqr[\Ueps] u + \normsqr[\Ueps] {du}\right)^{1/2}  \\=
    O(\eps^{(\alpha d - m)/2})(\normsqr[\Ueps \cup \Veps] u +
    \normsqr[\Ueps \cup \Veps]
    {du})^{1/2}
  \end{multline*}
  % ------------- %
  where we have used Lemmas~\ref{lem:diff.vol.av},~\ref{lem:diff.av},
  and~\ref{lem:av.edge}, and furthermore the
  inequality~\eqref{eq:est.vert.norm} to obtain $\vol \Veps = O(\eps^{\alpha
    d})$.  \sloppy Finally, note that $\beta = (d+2)\alpha - d > \alpha d - m
  >0$ and that $\alpha d - m > 0$ is equivalent to
  assumption~\eqref{eq:est.alpha.fast}.
\end{proof}
% ------------- %

Now we define the transition operator by
% ------------- %
\begin{equation}
  \label{eq:trans.op.below}
  (\Psi_\eps u)_j (x):=
    \eps^{m/2} (N_j u (x) + \rho(x)(C_k u - N_j u(x^0))
    \quad \text{for $x \in \Ijk$}
\end{equation}
% ------------- %
where $\map \rho \R {[0,1]}$ is a smooth function such that
% ------------- %
\begin{equation}
  \label{eq:smooth.fct}
  \rho(x^0)=1           \quad \text{and} \quad
  \rho(x)=0
      \quad \text{for all $|x-x^0| \ge \frac12 \min_{j \in J} \ell_j$}
\end{equation}
% ------------- %
where $\ell_j$ denotes the length of the edge $e_j \cong \Ij$.  Furthermore,
$x^0=x^0_{jk} \in \bd \Ij$ is the edge point which can be identified with the
vertex $v_k$. Recall that $\Ijk$ denotes the (closed) half of the interval
$\Ij \cong e_j$ adjacent with the vertex $v_k$ and directed away from $v_k$.

% ------------- %
\begin{lem}
  \label{lem:below}
  We have $\Psi_\eps u \in \Sob \Mnull$ if $u \in \Sob \Meps$.
  Furthermore,
  % ------------- %
  \begin{align}
    \normsqr[\Meps] u - \normsqr[\Mnull] {\Psi_\eps u} &\le
    o(1) (\normsqr[\Meps] u + \normsqr[\Meps] {du}) \\
    q_0(\Psi_\eps u) - \normsqr[\Meps] {du} &\le
    o(1) (\normsqr[\Meps] u + \normsqr[\Meps] {du})
  \end{align}
  % ------------- %
  for all $u \in \Sob \Meps$.
\end{lem}
% ------------- %
\begin{proof}
  The first assertion follows from $(\Psi_\eps u)_j(x^0_{jk}) = C_k u$.
  Furthermore, we have
  % ------------- %
  \begin{multline*}
    \normsqr[\Meps] u - \normsqr[\Mnull] {\Psi_\eps u} \\ \le
    \sum_{k \in K}
    \Bigl(
      \normsqr[\Vepsk] u +
      \sum_{j \in J_k}
      \bigl( 
          \normsqr[\Uepsjk] u - \eps^m \normsqr[\Ijk]
        {Nu + \rho \cdot (Cu - Nu(x^0))}
      \bigr)
    \Bigr) \\ \le
    \sum_{k \in K}
    \Bigl(
      \normsqr[\Vepsk] u +
      \sum_{j \in J_k}
      \bigl(
        \normsqr[\Uepsjk] u - \normsqr[\Ijk] {\eps^{m/2} Nu}
      \bigr)
          \\ +
      \sum_{j \in J_k}
      \bigl(
        \delta \normsqr[\Ijk] {\eps^{m/2} Nu} +  \eps^m \delta^{-1}
       \normsqr[\Ijk] \rho |Cu - Nu(x^0)|^2
      \bigr)
     \Bigr)
  \end{multline*}
  % ------------- %
  where we have used the inequality
  % ------------- %
  \begin{equation}
    \label{eq:quad.lower}
    (a+b)^2 \ge (1-\delta)a^2 - \frac 1 \delta b^2, \qquad \delta>0.
  \end{equation}
  % ------------- %
  The last term in the sum can be estimated by
  $O(\eps^m)\delta^{-1}|C u - N  u(x^0)|^2$. Applying
  Lemma~\ref{lem:diff.av} we arrive at the bound by
  $O(\eps^{m+2\alpha-d})\delta^{-1}(\normsqr[\Meps] u + \normsqr[\Meps] {du})$.
  Note that $m+2\alpha-d=2\alpha-1>0$ since $\alpha>1/2$. Set
  $\delta := \eps^{(2\alpha-1)/2}$. The remaining terms can be estimated
  by Corollary~\ref{cor:vertex.small},
  Lemma~\ref{lem:diff.norm.av}, and estimate~\eqref{eq:edge.av.cont}.

  The second inequality can be proven in the same way, namely
  % ------------- %
  \begin{multline*}
    q_0(\Psi_\eps u) - \normsqr[\Meps] {du}  \le
    \sum_{\substack{k \in K\\ j \in J_k}}
    \Bigl(
      \eps^m \normsqr[\Ijk]
        {(Nu)' + \rho' \cdot (Cu - Nu(x^0))} -
        \normsqr[\Uepsjk] {du}
    \Bigr) \\ \le
    \sum_{\substack{k \in K\\ j \in J_k}}
    \bigl(
       \normsqr[\Ijk] {\eps^{m/2} (Nu)'} - \normsqr[\Uepsjk] {du}
     +
       \delta \normsqr[\Ijk]  {\eps^{m/2} (Nu)'}
       + \frac{2\eps^m} \delta \normsqr[\Ijk] {\rho'}
         |Cu - Nu(x^0)|^2
    \bigr)
  \end{multline*}
  % ------------- %
  where we have used
  % ------------- %
  \begin{equation}
    \label{eq:quad.upper}
    (a+b)^2 \le (1+\delta)a^2 + \frac 2 \delta b^2, \qquad 0 < \delta \le 1,
  \end{equation}
  % ------------- %
  with $\delta := \eps^{(2\alpha-1)/2}$. Since the norm involving
  $\rho'$ is a fixed constant, the result follows from
  Lemma~\ref{lem:diff.quad.av} and Lemma~\ref{lem:diff.av}.
\end{proof}

Using Lemma~\ref{lem:below} we arrive at the sought lower bound.
Note that the error term $\eta_k$ in~\eqref{eq:eta.k} can be
estimated by some $\eps$-independent quantity because
$\lambda_k=\EW k \Meps \le c_k$ by the upper bound given in
Theorem~\ref{thm:ev.above}.
% ------------- %
\begin{thm}
\label{thm:ev.below}
  We have $\EW k \Mnull \le \EW k \Meps + o(1)$.
\end{thm}
% ------------- %
\noindent Theorem~\ref{thm:ev.conv} now follows easily by
combining the last result with Theorem~\ref{thm:ev.above}.

%------------------------------------------------------------
\section{Slowly decaying vertex volume}
\label{sec:slow.decay}
%------------------------------------------------------------
If the volume of the vertex region decays significantly slower
than the volume of the edge neighbourhoods, the limit operator is
different. At the ends of the edges we have Dirichlet boundary
conditions, whereas for each vertex $v_k$, $k \in K$, we obtain an
additional eigenmode. In other words, we add a point measure at
each vertex to the given measure on the graph $\Mnull$; the
corresponding Hilbert space and quadratic form (domain) is
therefore given by
% ------------- %
\begin{equation}
  \label{def:lim.slow}
  \HS_0 := \Lsqr \Mnull \oplus \C^K, \qquad
  \mathcal D_0 := \bigoplus_j \Sobn \Ij \oplus \C^K, \qquad
  q_0(u):= \sum_j \normsqr[\Ij] {u_j'}.
\end{equation}
% ------------- %
For elements of $\HS_0$ we write $u=((u_j)_{j \in J},(u_k)_{k \in K})$ where
$u_j \in \Lsqr {\Ij, p_j(x)dx}$ and $u_k \in \C$. We sometimes omit the
indices and simply write $u$ instead of $u_j$. Note that the point
contributions $u_k$ do not occur in the quadratic form, i.e., the additional
eigenmodes have zero energy. Furthermore, the associated operator
\begin{displaymath}
  Q_0 := \bigoplus_{j \in J} \laplacianD \Ij \oplus
         \boldsymbol 0
\end{displaymath}
corresponds to a fully decoupled graph, i.e., a collection of independent
edges, and its spectrum consists of all Dirichlet eigenvalues of the intervals
$\Ij$ and $0$. Here, $\boldsymbol 0$ corresponds to the zero operator on
$\C^K$.

In order to define assumptions such that a smooth metric $g_\eps$ exists
globally with different length scalings on the vertex and edge neighbourhoods,
we need to introduce some additional notation (see Figure~\ref{fig:add.mfd}):
% ------------- %
\begin{figure}[h]
  \begin{center}
%------------------------------------------------------------
%\newcommand{\color}[2][{}]{}
%    \input{graph3.pstex_t}
\begin{picture}(0,0)%
  \includegraphics{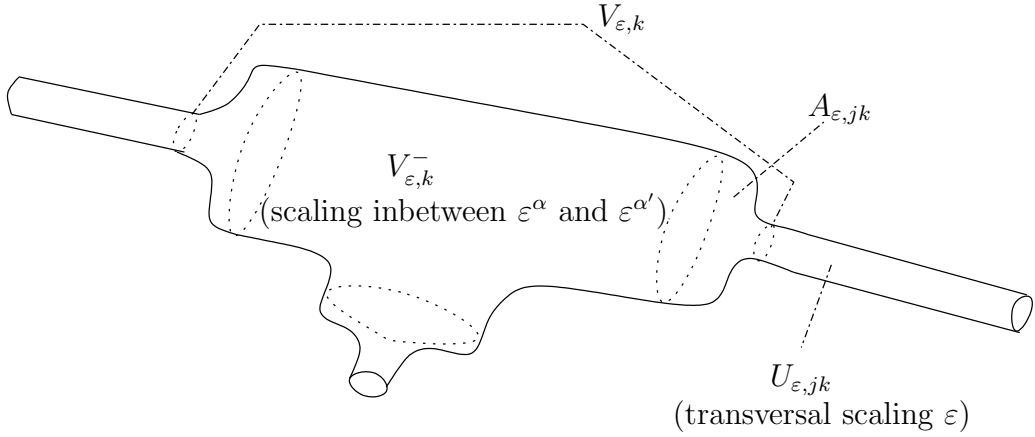}%
\end{picture}%
\setlength{\unitlength}{4144sp}%
\begin{picture}(6155,2598)(55,-1963)
  \put(4861,-61){$\Aepsjk$}%
  \put(3601,479){$\Vepsk$}%
  \put(4636,-1681){$\Uepsjk$}%
  \put(2341,-421){$\mVepsk$}%
  \put(1576,-691){(scaling inbetween $\eps^\alpha$ and $\eps^{\alpha'}$)}%
  \put(4052,-1905){(transversal scaling $\eps$)}%
\end{picture}
%------------------------------------------------------------
    \caption{The decomposition with the different scaling areas.}
    \label{fig:add.mfd}
  \end{center}
\end{figure}
% ------------- %
let $\mVk$ be a closed submanifold of $\Vk$ of the same dimension with a
positive distance from all adjacent edge neighbourhoods $\Ujk$, $j \in J_k$.
Furthermore, we assume that the cylindrical structure of the half vertex
neighbourhood $\Ujk$ extends to the component of $\Vk \setminus \mVk$ where
$\Ujk$ meets $\Vk$, i.e., the closure of $\Vk \setminus \mVk$ is diffeomorphic
to the disjoint union of cylinders $[0,1] \times F$. We denote the extended
cylinder containing $\Ujk$ together with the corresponding cylindrical end
(the \emph{bottle neck}) of $\Vk$ by $\pUjk=\pIjk \times F$ and the bottle neck
alone by $\Ajk = \nIjk \times F$.  Note that $\Ajk = \pUjk \cap \Vk$ and that
$\pIjk = \Ij \cup \nIjk$.

Again, we use the subscript $\eps$ to indicate the corresponding
Riemannian manifold with metric $g_\eps$.

%------------------------------------------------------------
\subsection{Assumption on the smaller vertex neighbourhood}
We first fix the scaling behaviour on the smaller vertex neighbourhood
$\mVk$. Here, we assume that
% ------------- %
\begin{equation}
  \label{eq:met.vertex.slow}
  c_- \eps^{2\alpha} g \le g_\eps \le c_+ \eps^{2\alpha'} g
  \qquad \text{on $\mVk$}
\end{equation}
(for the notation see~\eqref{eq:met.vertex}) where
\begin{equation}
  \label{eq:est.alpha.slow}
  0 < \alpha < \frac{d-1} d,
\end{equation}
% ------------- %
i.e., $\mVepsk$ scales at most as $\eps^\alpha$ in each direction and at least
as $\eps^{\alpha'}$ where
% ------------- %
 \begin{equation}
  \label{eq:est.alpha'.slow}
  \frac d {d+2} \alpha  < \alpha' \le \alpha,
\end{equation}
e.g., a homogeneous scaling ($\alpha'=\alpha$) would do. Note that $\alpha'
\le \alpha$ is necessary in order that \eqref{eq:met.vertex.slow} makes sense
whereas $\alpha d/(d+2) < \alpha'$ ensures that the second Neumann eigenvalue
of $\mVeps$ tends to $\infty$ as we will need in
Lemma~\ref{lem:diff.vol.av.slow}.

%------------------------------------------------------------
\subsection{Assumptions on the bottle neck}
Roughly speaking, we have to avoid that the bottle neck has more than a single
neck separating $\Vepsk$ in more than one part as $\eps \to 0$. In that case
more than one zero eigenmode occur in the limit.

We use the same notation as in Section~\ref{sec:edge.nbh} for the metric
$g_\eps$ on the bottle neck $A=\Ajk$ and set
\begin{equation}
  \label{eq:def.met.edge.slow}
  \tilde g_\eps := a_\eps^2(x) dx^2 + r_\eps^2(x) h(y), \qquad 
    (x,y) \in A=\nI \times F
\end{equation}
for the (pure) product metric on $A$. Here, $a_\eps=a_{\eps,jk}$ and
$r_\eps=r_{\eps,jk}$ are strictly positive smooth functions. Note that $r_\eps$
defines the radius of the fibre $\{x\} \times F$ at the point $x$. Again, we
denote by $G_\eps$ and $\tilde G_\eps$ the $d \times d$-matrices associated to
the metrics $g_\eps$ and $\tilde g_\eps$ with respect to the coordinates
$(x,y) \in \nI \times F$) and assume that the two metrics coincide up to an
error term as $\eps\to 0$, more specifically
% ------------- %
\begin{equation}
  \label{eq:asym.met.edge.slow}
  G_\eps = \tilde G_\eps +
    \begin{pmatrix}
      o(a_\eps^2) & o(a_\eps r_\eps) \\ o(a_\eps r_\eps) &
      o(r_\eps^2)
    \end{pmatrix} =
    \begin{pmatrix}
      (1+o(1)) a_\eps^2  & o(a_\eps r_\eps) \\ o(a_\eps r_\eps) &
      (H + o(1)) r_\eps^2
    \end{pmatrix},
\end{equation}
% ------------- %
uniformly on $A$.

We prove the following lemma in the same way as Lemma~\ref{lem:metric}:
\begin{lem}
  \label{lem:metric.slow}
  Suppose that $g_\eps$, $\tilde g_\eps$ are given as above then
  % ------------- %
  \begin{align}
  \label{eq:met.vol.slow}
    (\det G_\eps)^\frac12 &= (1 + o(1)) \, (\det \tilde G_\eps)^\frac12 \\
  \label{eq:met.1st.comp.slow}
    g_\eps^{xx} & \!:= (G_\eps^{-1})_{xx} = a_\eps^{-2}(1 + o(1)) \\
  \label{eq:met.1st.der.slow}
    a_\eps^{-2} |d_x u|^2 &\le O(1) \, |d u|_{g_\eps}^2
  \end{align}
  % ------------- %
  where $d_x$ denotes the partial derivative with respect to $x$.
\end{lem}

To make a smooth junction between the metrics on $\Uj$ and $\mVk$ possible, we
assume that
\begin{align*}
  a_\eps(x)&=\eps^\alpha, &  r_\eps(x)&=\eps^\alpha     &&\text{near $x^+$}\\
  a_\eps(x)&=1,           &  r_\eps(x)&=\eps r_-        &&\text{near $x^0$}
\end{align*}
where $x \in \nI=[x^+,x^0]$ and $r_-:=r_j(x^0)$ (the radius of the fibre at
$x^0$, see also equation~\eqref{eq:def.met.edge}).

Furthermore, we assume that
\begin{equation}
  \label{eq:est.a.r}
  a_\eps(x) \le
  \begin{cases}
    \eps^\alpha & \text{on $[x^+, x^0-\delta_0,]$}\\
    1 & \text{on $[x^0-\delta_0, x^0]$}
  \end{cases}
  \quad
  \eps r_- \le r_\eps(x) \le 
  \begin{cases}
    \eps^\alpha  & \text{on $[ x^+, x^+ + \delta_+]$}\\
    \eps r_+ & \text{on $[x^+ + \delta_+, x^0]$}
  \end{cases}
\end{equation}
for some constant $r_+ \ge r_-$, where $\delta_0=\eps^\alpha$ and
\begin{figure}[h]
  \begin{center}
%------------------------------------------------------------
%    \input{a.r.pstex_t}
\begin{picture}(0,0)%
  \includegraphics{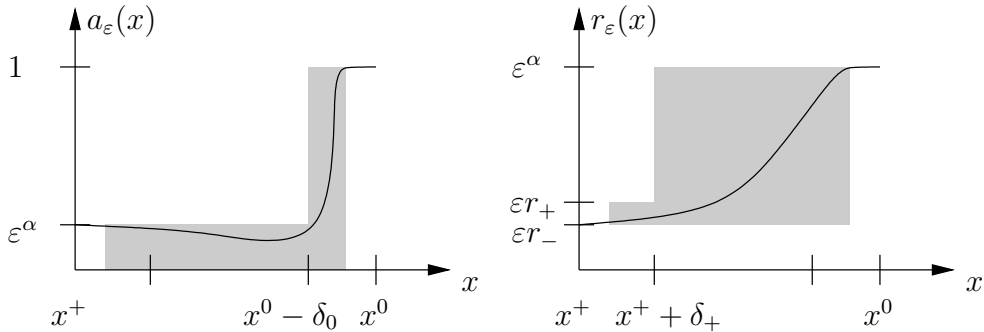}%
\end{picture}%
\setlength{\unitlength}{4144sp}%
\begin{picture}(5730,1926)(58,-1276)
  \put(530,494){$a_\eps(x)$}%
  \put(2773,-1073){$x$}%
  \put( 58,209){$1$}%
  \put( 58,-781){$\eps^\alpha$}%
  \put(3545,494){$r_\eps(x)$}%
  \put(5788,-1073){$x$}%
  \put(3073,209){$\eps^\alpha$}%
  \put(3073,-781){$\eps r_-$}%
  \put(3061,-601){$\eps r_+$}%
  \put(1466,-1276){$x^0-\delta_0$}%1666
  \put(2161,-1276){$x^0$}%
  \put(316,-1276){$x^+$}%
  \put(3691,-1276){$x^++\delta_+$}%
  \put(5176,-1276){$x^0$}%
  \put(3331,-1276){$x^+$}%
\end{picture}
%------------------------------------------------------------
    \caption{The functions $a_\eps$ and $r_\eps$ in its allowed range (in
      grey).}
    \label{fig:a.r}
  \end{center}
\end{figure}
$\delta_+=\eps^{(1-\alpha)m}=\eps^\alpha \eps^{m - \alpha d}$. These
assumptions are needed in Lemma~\ref{lem:poincare}, e.g.\ to assure that the
eigenfunctions of $\Meps$ do not concentrate on $\Aepsjk$ (i.e.,
\eqref{eq:rest.small} holds).

%------------------------------------------------------------
\subsection{Convergence of the spectra}

With the above prerequisites we can finally formulate the main
result of this section:
% ------------- %
\begin{thm}
  \label{thm:ev.conv.slow}
  Under the stated assumptions $\EW k \Meps \to \EW k {Q_0}$ as $\eps\to 0$.
  More precisely, the first $|K|$ eigenvalues tend to $0$, while the remaining
  bounded eigenvalue branches tend to Dirichlet eigenvalues of the intervals
  $I_j$, i.e.,
  \begin{align}
    \label{eq:lim.slow.0}
     \EW k \Meps &\to 0 & \text{if $1 \le k \le |K|$}\\
    \label{eq:lim.slow.bigger.0}
     \EW k \Meps &\to \EWD {k-|K|}
  {\bigdcup_{j \in J} I_j} & \text{if $k > |K|$},
\intertext{where $\EWD n {\bigdcup_{j \in J} I_j}$ denotes the Dirichlet
  eigenvalues $\EWD l {\laplacian \Ij}$ of the operators on $\Ij$ ($j \in J$)
  defined as in~\eqref{eq:formaledge}, reordered with respect to
  multiplicity. In particular, if the length of all the edges $I_j$ is $\ell$
  and $p_j(x)=1$ for all $j$, we have}
    \EW k \Meps &\to \EWD m {[0,\ell]} = \pi^2 m^2/\ell^2 & \text{if
     $k=(m-1)|J|+1,\dots, m|J|$.}
  \end{align}
\end{thm}
% ------------- %

Again, our aim is to find a two sided estimate on each eigenvalue $\EW k
\Meps$ by means of $\EW k {Q_0}$ with an error which is $o(1)$
w.r.t.\ the parameter $\eps$.

%------------------------------------------------------------
\subsection{An upper bound}
%\label{ssec:up.est}

The following upper eigenvalue estimate is slightly more difficult to show
than in the previous section:
% ------------- %
\begin{thm}
\label{thm:ev.above.slow}
  $\EW k \Meps \le \EW k {Q_0} + o(1)$ holds as $\eps\to 0$.
\end{thm}
% ------------- %
\noindent To prove it, we define the transition operator by
% ------------- %
\begin{displaymath}
  \label{eq:trans.op.above.slow}
  \Phi_\eps u (z):=
  \begin{cases}
    (\vol \mVepsk)^{-1/2} u_k & \text{if $z \in \Vk$},\\
    \eps^{-m/2} u_j(x) + (\vol \mVepsk)^{-1/2} \rho(x) u_k &
           \text{if $z=(x,y) \in \Uj$}
  \end{cases}
\end{displaymath}
% ------------- %
for any $u \in \mathcal D_0$, where $\rho$ is a smooth function as
in~\eqref{eq:smooth.fct} and $x^0=x^0_{jk}$ denotes the endpoint of
the half-edge $\Ijk$ corresponding to the vertex $v_k$.
Theorem~\ref{thm:ev.above.slow} is then implied by
Lemma~\ref{lem:main} in combination with the following result.

% ------------- %
\begin{lem}
  \label{lem:above.slow}
  We have $\Phi_\eps u \in \Sob \Meps$, i.e., $\Phi_\eps$ maps the quadratic
  form domain $\mathcal D_0$ into the quadratic form domain of the Laplacian
  on the manifold.  Furthermore,
% ------------- %
  \begin{align}
    \normsqr[\HS_0] u - \normsqr[\Meps] {\Phi_\eps u} &\le
    o(1) \, \normsqr[\HS_0] u\\
    \normsqr[\Meps] {d \, \Phi_\eps u} - q_0(u) & \le
    o(1) \, (\normsqr[\HS_0] u + q_0(u))
  \end{align}
% ------------- %
\end{lem}
% ------------- %
\begin{proof}
  Since $u_j \restr {\bd \Ij} = 0$, the function $\Phi_\eps u$ agrees
  on $\bd \Vepsk$ for both definitions. Clearly, $\Phi_\eps u$ is
  weakly differentiable on each thickened edge $\Uepsj$.  Moreover, we
  have
  % ------------- %
  \begin{multline*}
    \normsqr[\HS_0] u - \normsqr[\Meps] {\Phi_\eps u}
\le
    \sum_{\substack{k \in K\\ j \in J_k}}
%    \sum_{k \in K, j \in J_k}
    \Bigl(
      \bigl(
         \normsqr[\Ijk] u -
         (1+o(1)) \Bignormsqr[\tUepsjk]
         {\eps^{-\frac m2} u + (\vol \mVepsk)^{-\frac12} \rho \, u_k}
      \bigr)
+{}\\{}+
%    \right) \\
%    + \sum_{k \in K}
%        \left(
         \bigl(   
            |u_k|^2 - \normsqr[\mVepsk] {\Phi_\eps u}
         \bigr)
   \Bigr)
  \end{multline*}
  % ------------- %
  where we have used equation~\eqref{eq:met.vol} and that $\mVepsk \subset
  \Vepsk$. Note that the latter sum in the last line is equal to $0$. To
  estimate the remaining sum, remember that $\Phi_\eps u$ is independent of
  $y$ on $\Uepsjk$. Therefore we can apply equation~\eqref{eq:ind.of.2nd}, and
  inequality~\eqref{eq:quad.lower} with $\delta:=\eps^{(m-\alpha d)/2}$ yields
  the upper estimate
  % ------------- %
  \begin{displaymath}
%     \sum_{\substack{k \in K\\ j \in J_k}}
%     \biggl(
%       \normsqr[\Ijk] u -
%       (1+o(1))
%       \Bigl(
%          (1-\delta)
%       \normsqr[\Ijk] u  - 
%       \frac{\eps^m}{\delta \vol \mVepsk}
%          \normsqr[\Ijk]\rho |u_k|^2
%       \Bigr)
%     \biggr) \\ \le
%    \sum_{\substack{k \in K\\ j \in J_k}}
    \sum_{k \in K, j \in J_k}
    \Bigl(
      (\delta + o(1)) \normsqr[\Ijk] u +
      \frac{O(\eps^{m-\alpha d})}{\delta} |u_k|^2
    \Bigr) =
      o(1) \normsqr[\HS_0] u.
   \end{displaymath}
  % ------------- %
   In the last inequality, we have used the estimate $(\vol \mVepsk)^{-1} \le
   O(\eps^{-\alpha d})$ which follows from the lower bound
   of~\eqref{eq:met.vertex.slow}. Note that $\delta=o(1)$ by
   assumption~\eqref{eq:est.alpha.slow}.  The second relation follows from
  % ------------- %
  \begin{multline*}
    \normsqr[\Meps] {d \, \Phi_\eps u} - q_0(u) =
%     \sum_{\substack{k \in K\\ j \in J_k}}
%     \biggl(
%       \bigl(1+o(1)\bigr) \normsqr[\tUepsjk]{g_\eps^{xx} d_x \Phi_\eps u} -
%              \normsqr[\Uepsjk]{u'}
%     \biggr)\\ =
    \sum_{\substack{k \in K\\ j \in J_k}}
    \biggl(
        \bigl(1+o(1)\bigr) \eps^m \normsqr[\Ijk]{\eps^{-\frac m2}u' +
                  (\vol \mVepsk)^{-\frac12} \rho'\, u_k} -
                 \normsqr[\Ijk]{u'}
    \biggr) \\ \le
    \sum_{\substack{k \in K\\ j \in J_k}}
    \bigl(1+o(1)\bigr)
    \biggl(
      \delta \normsqr[\Ijk]{u'} +
      \frac{2 \eps^m}{\delta \vol \mVepsk}
          \normsqr[\Ijk] {\rho'} |u_k|^2 
    \biggr) \le
     o(1) \bigl(\normsqr[\HS_0] u + q_0(u)\bigr)
  \end{multline*}
  % ------------- %
  in the same way as above together with \eqref{eq:met.1st.comp} for the
  second equality and~\eqref{eq:quad.upper} in the last line; recall that
  $\Phi_\eps u$ is constant on $\Vepsk$.
\end{proof}

Note that we need a counterpart to $\normsqr[\Vepsk] u$ on the limit Hilbert
space $\HS_0$. In the case of a fast decaying vertex volume in the previous
section, the corresponding norm vanished (see
Corollary~\ref{cor:vertex.small}), but here we need the additional subspace
$\C^K$ in $\HS_0$ coming from extra point measures at the vertices.

Furthermore, note that the upper bound estimate on $\EW k \Meps$ already
proven in Lemma~\ref{lem:above} remains valid in this setting, but it is too
rough for the present purpose.

\subsection{A lower bound}
%\label{ssec:low.est}

Again, the reverse estimate is more difficult. We will employ
averaging processes also on the vertex neighbourhoods; this time with
the $\eps$-scaled manifold $\mVepsk$:
% ------------- %
\begin{equation}
  \label{def:vert.av.eps}
  C_\eps^- u = C_{\eps,k}^- u :=
  \frac 1 {\vol {\mVepsk}} \int_{\mVepsk} u \, d \mVepsk.
\end{equation}
% ------------- %
In the first lemma, we prove an estimate similar to the one in
Lemma~\ref{lem:diff.vol.av}. Note that $\norm {C_\eps^- u} \le \norm u$ by
Cauchy-Schwarz, but we need the reverse inequality.
% ------------- %
\begin{lem}
  \label{lem:diff.vol.av.slow}
  For any $u \in \Sob \mVeps$ we have
% ------------- %
  \begin{displaymath}
    \normsqr[\mVeps] u - \normsqr[\mVeps] {C_\eps^- u} \le
    o(1) % O(\eps^{((d+2)\alpha' - d \alpha)/2})
     \bigl(\normsqr[\mVeps] u + \normsqr[\mVeps] {du} \bigr).
  \end{displaymath}
% ------------- %
\end{lem}
% ------------- %
\begin{proof}
  Apply Lemma~\ref{lem:minmax.2nd.neu} with $X=\mVeps$ and $\delta
  =\eps^{((d+2)\alpha' - d \alpha)/2}$. From the min-max principle we obtain
  $\EWN 2 \mVeps \ge O(\eps^{d(\alpha-\alpha')-2\alpha'})$. Note that
  $\delta=o(1)$ since $(d+2)\alpha' - d \alpha>0$
  by~\eqref{eq:est.alpha'.slow}.
\end{proof}
% ------------- %

The next three results are valid independently of the assumptions on $\alpha$
given in~\eqref{eq:est.alpha.slow} and \eqref{eq:est.alpha'.slow}. We will
need these results also for the borderline case $\alpha=(d-1)/d$ in the next
section.

We need an estimate on the average $|Nu(x^0)|^2$.  Since on the bottle neck
$\Aepsjk$, the estimates are quite delicate, we first prove the result for
$|Nu(x^+)|^2$ (i.e., on $\bd \mVepsk$ where the scaling of the metric is of the
right order. The error is controlled by~\eqref{eq:diff.av.slow}. Note that
this estimate is a counterpart to the estimate in Lemma~\ref{lem:av.edge}
where we extended the function to the \emph{edge} neighbourhood $\Uepsj$
(useful in the case of fast decaying vertex volume, $\alpha d - m > 0$). This
is not possible here, since $\alpha d - m < 0$. Therefore, we extend the
function to the \emph{vertex} neighbourhood $\mVepsk$.

% ------------- %
\begin{lem}
  \label{lem:av.vert.slow}
  The inequality
% ------------- %
  \begin{displaymath}
    |Nu(x^+)|^2 \le \normsqr[F] {u(x^+,\cdot)} \le
    O(\eps^{-\alpha d}) \,
    \bigl( \normsqr[\mVeps] u + \normsqr[\mVeps]{du} \bigr)
  \end{displaymath}
% ------------- %
  holds for any $u \in \Sob \mVeps$.
\end{lem}
% ------------- %

\begin{proof}
  We have
% ------------- %
  \begin{multline*}
    |Nu(x^+)|^2 \le \int_F |u(x^+,y)|^2 dF(y) \\ \le
    c_1 \bigl( \normsqr[\mV] u + \normsqr[\mV]{du} \bigr) \le
    O(\eps^{-\alpha d}) \bigl( \normsqr[\mVeps] u + \normsqr[\mVeps]{du} \bigr)
  \end{multline*}
% ------------- %
  by Lemma~\ref{lem:rest.est} with $X=\mV$ and the lower bound in
  assumption~\eqref{eq:est.alpha.slow}.
\end{proof}
% ------------- %

The next lemma is the key ingredient in dealing with the bottle neck.  Here,
we prove two Poincar\'e-like estimates. Since we want to avoid a cut-off
function (leading to divergent terms when being differentiated) we only prove
an estimate on the \emph{difference} and not on $Nu(x^0)$ itself
in~\eqref{eq:diff.av.slow}. For the same reason, an integral over $F$ remains
in~\eqref{eq:rest.small}. Note that $I_\eps^+(x^0) = \vol \Aeps$.

% ------------- %
\begin{lem}
\label{lem:poincare}
There is a constant $C>0$ such that
% ------------- %
\begin{gather}
  \label{eq:diff.av.slow}
  |Nu(x^0) - Nu(x^+)|^2 \le
  C  I_\eps^-(x^0) \, \normsqr[\Aeps] {du},  \\
  \label{eq:rest.small}
  \normsqr[\Aeps] u  \le
     4 I_\eps^+(x^0) \, \normsqr[F] {u(x^+,\cdot)} +
     4 C I_\eps^{+-} \, \normsqr[\Aeps]{du}
\end{gather}
% ------------- %
for all $u \in \Sob \Aeps$ where
\begin{displaymath}
  I_\eps^\pm(x) := \int_{x+}^x     a_\eps(x') r_\eps^{\pm m}(x') \, dx'
  \quad \text{and} \quad
  I_\eps^{+-} := \int_{x^+}^{x^0} a_\eps(x)  r_\eps^m(x) I_\eps^-(x) \, dx.
\end{displaymath}
Furthermore, under the assumption~\eqref{eq:est.a.r}, we have
\begin{displaymath}
    I_\eps^-(x^0) = o(\eps^{-m}), \qquad
    I_\eps^+(x^0) = o(\eps^{\alpha d}) \qquad \text{and} \qquad
    I_\eps^{+-} = o(1).
\end{displaymath}
% ------------- %
\end{lem}

% ------------- %

\begin{proof}
  For a smooth function $u$ we have
% ------------- %
  \begin{equation}
    \label{eq:diff.int}
    u(x,y) - u(x^+,y) =  \int_{x^+}^x \partial_x u(x',y) \, dx'.
% ------------- %
  \end{equation}
  For the first assertion, we
  set $x=x^0$, foremost integrate over $y \in F$ and then apply
  Cauchy-Schwarz
% ------------- %
  \begin{multline*}
    |Nu(x^0) - Nu(x^+)|^2  \le
    \int_F \int_{x^+}^{x^0} a_\eps^2(x') \det G_\eps(x',y)^{-\frac12} \, dx'
    \times{} \\ {} \times
    \int_{x^+}^{x^0}  a_\eps^{-2}(x') |\partial_x u(x',y)|^2 \det
             G_\eps(x',y)^\frac12 \, dx'\, dF(y).
  \end{multline*}
% ------------- %
  The first integrand over $x'$ can be estimated by $C a_\eps(x') r_\eps^{-m}$
  applying~\eqref{eq:met.vol.slow}. Therefore, the first integral is smaller
  than $C I_\eps^-(x^0)$. The second integral together with the integral over
  $F$ can be estimated by $O(1) \, \normsqr[\Aeps]{du}$
  applying~\eqref{eq:met.1st.der.slow}.

  For the second assertion, we first apply Cauchy-Schwarz
  (and~\eqref{eq:quad.upper} with $\delta=1$) to~\eqref{eq:diff.int} and than
  integrate over $y \in F$ to obtain
% ------------- %
  \begin{multline}
    \label{eq:rest.est}
    \int_F |u(x,y)|^2 \, dF(y) \le
    2 \int_F |u(x^+,y)|^2 \, dF(y)  +
    2 \int_F \int_{x^+}^x a_\eps^2(x') \det G(x',y)^{-\frac12} \, dx' 
      \times {}\\ {} \times
     \int_{x^+}^x  a_\eps^{-2} (x') \, |\partial_x u (x',y)|^2  
        \det G(x',y)^\frac12  \,dx' \,dF(y).
  \end{multline}
% ------------- %
  The first integral over $x'$ can be estimated as before by $C 
  I_\eps^-(x)$. Finally, multiplying with $a_\eps(x) r_\eps^m(x)$ and
  integrating over $x \in \nI$ yields
  \begin{displaymath}
    \normsqr[\tAeps] u \le
    2 I^+_\eps(x_0) \, \normsqr[F]{u(x^+,\cdot)} + 
    2 C  I_\eps^{+-} \,\normsqr[\Aeps]{du}.
  \end{displaymath}
  Applying~\eqref{eq:met.vol.slow} once more we obtain the desired estimate
  over $\Aeps$ instead of $\tAeps$ (note that $2/(1+o(1)) \le 4$ provided
  $\eps$ is small enough).  The general case of non-smooth functions can
  easily shown with approximation arguments.
  
  The integral estimates follow from
  \begin{displaymath}
    I_\eps^-(x^0) \le
    \int_{x^+}^{x^0 - \delta_0} \eps^\alpha (\eps r_-)^{-m} \, dx +
    \int_{x^0-\delta_0}^{x^0} (\eps r_-)^{-m} \, dx  =
    (\eps^\alpha + \delta_0) O(\eps^{-m}).
  \end{displaymath}
  Since $\delta_0=\eps^\alpha$, we have
  $I_\eps^-(x^0) \le O(\eps^{\alpha-m})$. Next, we have
  \begin{multline*}
    I_\eps^+(x^0) \le
     \int_{x^+}^{x^+ + \delta_+} \eps^\alpha \eps^{\alpha m} \, dx +
     \int_{x^+ + \delta_+}^{x^0 - \delta_0} \eps^\alpha (\eps r_+)^m \, dx \\ +
     \int_{x^0-\delta_0}^{x^0} (\eps r_+)^m \, dx=
     \delta_+ O(\eps^{\alpha d}) + ( \eps^\alpha + \delta_0)  O(\eps^m)   
  \end{multline*}
  and therefore $I_\eps^+(x^0) \le O(\eps^{\alpha + m})= O(\eps^{\alpha + m -
  \alpha d}) O(\eps^{\alpha d})$ since $\delta_+=\eps^\alpha \eps^{m-\alpha
  d}$.
  The last assertion follows from $I_\eps^{+-} \le I_\eps^-(x_0) \,
  I_\eps^+(x_0) \le O(\eps^{2\alpha})$.
\end{proof}

The following corollary is again independent of the assumption we made about
$\alpha$ in~\eqref{eq:est.alpha.slow} and \eqref{eq:est.alpha'.slow}, in
particular, it is also valid in the setting of the borderline case of
Section~\ref{sec:borderline}.
% ------------- %
\begin{cor}
  \label{cor:rest.small}
  For all $u \in \Sob \Veps$ we have
% ------------- %
  \begin{displaymath}
    \normsqr[\Aeps] u \le o(1) \,
    \bigl( \normsqr[\Veps] u + \normsqr[\Veps]{du} \bigr).
  \end{displaymath}
% ------------- %
\end{cor}
% ------------- %
\begin{proof}
  We only have to put together~\eqref{eq:rest.small} and
  Lemma~\ref{lem:av.vert.slow}.
\end{proof}
% ------------- %

We now formulate a consequence of the preceding lemmas under the
assumption~\eqref{eq:est.alpha.slow}.
\begin{cor}
  \label{cor:av.shifted}
  Suppose $0<\alpha< m/d=(d-1)/d$. Then we have
% ------------- %
  \begin{displaymath}
    |Nu(x^0)|^2 \le o(\eps^{-m}) \,
    \bigl( \normsqr[\Veps] u + \normsqr[\Veps] {du} \bigr)
  \end{displaymath}
% ------------- %
  for all $u \in \Sob \Veps$.
\end{cor}
% ------------- %
\begin{proof}
  Applying \eqref{eq:quad.lower} with $\delta=1/2$ to \eqref{eq:diff.av.slow}
  we obtain
% ------------- %
  \begin{displaymath}
    |Nu(x^0)|^2 \le
    o(\eps^{-m})  \normsqr[\Aeps]{du} +
    4|Nu(x^+)|^2.
  \end{displaymath}
% ------------- %
  The second term is of order $O(\eps^{-\alpha d})$ by
  Lemma~\ref{lem:av.vert.slow} and therefore also of order $o(\eps^{-m})$ by
  the assumption on $\alpha$.
\end{proof}
% ------------- %

In this section, we define the transition operator by
% ------------- %
\begin{equation}
  \label{eq:trans.op.below.slow}
  \begin{split}
    (\Psi_\eps u)_j (x) &:=
      \eps^{m/2} N_j u (x) -
                   \rho(x) N_j u (x^0)
      \quad \text{for $x \in I_{jk}$}\\
    (\Psi_\eps u)_k &:=
      (\vol \mVepsk)^{1/2} C_{\eps,k}^- u
  \end{split}
\end{equation}
% ------------- %
where $\rho$ is a smooth function as in~\eqref{eq:smooth.fct} and $x^0 =
x^0_{jk}$ denotes the endpoint of the half-edge $\Ijk$ corresponding to the
vertex $v_k$.
% ------------- %
\begin{lem}
  \label{lem:below.slow}
  We have $\Psi_\eps u \in \mathcal D_0$ if $u \in \Sob \Meps$.
  Furthermore,
  % ------------- %
  \begin{align}
    \normsqr[\Meps] u - \normsqr[\HS_0] {\Psi_\eps u} &\le
    o(1) \bigl(
       \normsqr[\Meps] u + \normsqr[\Meps] {du}
    \bigr) \\
    q_0(\Psi_\eps u) - \normsqr[\Meps] {du} &\le
    o(1) \bigl(
       \normsqr[\Meps] u + \normsqr[\Meps] {du}
    \bigr)
  \end{align}
  % ------------- %
  for all $u \in \Sob \Meps$.
\end{lem}
% ------------- %
\begin{proof}
  The first assertion follows from the fact that $(\Psi_\eps u)_j(x^0) = 0$.
  Furthermore, we have
  % ------------- %
  \begin{multline*}
    \normsqr[\Meps] u - \normsqr[\HS_0] {\Psi_\eps u}   \le
%     \sum_{k \in K}
%     \Bigr(
%        (\normsqr[\mVepsk] u - |(\Psi_\eps u)_k|^2) +
%        \sum_{j \in J_k}
%        \bigr(
%            \normsqr[\Aepsjk] u + \normsqr[\Uepsjk] u -
%            \normsqr[\Ijk] {(\Psi_\eps u)_j}
%        \bigr)
%     \Bigl) \\ \le
    \sum_{k \in K}
    \biggr(
       \bigl(
          \normsqr[\mVepsk] u - \normsqr[\mVepsk]{C_\eps^- u}
       \bigr) \\ +
       \sum_{j \in J_k}
       \bigr(
           \normsqr[\Aepsjk] u + \normsqr[\Uepsjk] u -
           \eps^m \bignormsqr[\Ijk] {N u -
                        \rho \cdot N u (x^0)}
       \bigr)
    \biggl).
  \end{multline*}
% ------------- %
  The first difference is of the desired form by
  Lemma~\ref{lem:diff.vol.av.slow}. Furthermore, the integral over the
  ``bottle necks'' $\Aepsjk$ can be estimated in the needed way by
  Corollary~\ref{cor:rest.small}.  Applying~\eqref{eq:quad.lower} to the
  remaining difference in the last sum we obtain the upper estimate by
% ------------- %
  \begin{equation}
    \label{eq:below.slow}
    \bigl( \normsqr[\Uepsjk] u - \eps^m \normsqr[\Ijk] {N u} \bigr) +
    \delta \, \eps^m \normsqr[\Ijk] {N u} +
    \frac {\eps^m} \delta \normsqr[\Ijk] \rho |N u (x^0)|^2
  \end{equation}
% ------------- %
  For the first two terms we obtain the sought bound by virtue of
  Lemma~\ref{lem:diff.norm.av} and \eqref{eq:edge.av.cont}; for the remaining
  term one has to apply Corollary~\ref{cor:av.shifted}.

  The second inequality can be proven in the same way, namely
   % ------------- %
   \begin{multline*}
     q_0(\Psi_\eps u) - \normsqr[\Meps] {du} \\ =
    \sum_{k \in K}
    \Bigr(
       -\normsqr[\mVepsk] {du} +
       \sum_{j \in J_k}
       \bigr(
           \eps^m \bignormsqr[\Ijk] {(N u)' -
                        \rho' \, N u (x^0)} -
           \normsqr[\Uepsjk] {du}
        \bigr)
    \Bigl)
   \end{multline*}
% ------------- %
   We omit the norm contribution from $\mVepsk$ and estimate the
   remaining difference with~\eqref{eq:quad.upper} and obtain (up to the
   summation)
% ------------- %
   \begin{displaymath}
    \bigl(\eps^m \normsqr[\Ijk] {(N u)'} - \normsqr[\Uepsjk] {du} \bigr) +
    \delta \, \eps^m \normsqr[\Ijk] {(N u)'} +
    2 \frac{\eps^m} \delta  \normsqr[\Ijk] {\rho'} |N u (x^0)|^2.
   \end{displaymath}
% ------------- %
   For the first difference we obtain the needed estimate by virtue of
   Lemma~\ref{lem:diff.quad.av}. An upper bound for the remaining term is of
   the same form as before.
\end{proof}
% ------------- %

Using Lemma~\ref{lem:below.slow} we arrive at the sought lower bound.
Note that the error term $\eta_k$ in~\eqref{eq:eta.k} can be
estimated by some $\eps$-independent quantity because
$\lambda_k=\EW k \Meps \le c_k$ by Theorem~\ref{thm:ev.above.slow}.
% ------------- %
\begin{thm}
\label{thm:ev.below.slow}
  We have $\EW k {Q_0} \le \EW k \Meps + o(1)$.
\end{thm}
% ------------- %
\noindent Theorem~\ref{thm:ev.conv.slow} now follows easily by
combining the last result with Theorem~\ref{thm:ev.above.slow}.

%------------------------------------------------------------
\section{The borderline case}
\label{sec:borderline}

% ------------- %
\subsection{Definition of the thickened vertices}
%\label{ssec:def.vert}
If the volume of the vertex region decays at the same rate as the
volume of the edge neighbourhoods, the limit operator acts again
in the extended Hilbert space introduced in the previous section
but it is not decoupled anymore. Thus it is not supported by the
graph alone, in particular, it is not the Hamiltonian with the
boundary conditions \eqref{eq:delta}.

We start with the definition of the limit operator.  The
corresponding Hilbert space and quadratic form are given by
% ------------- %
\begin{equation}
  \label{def:lim.border}
  \HS_0 := \Lsqr \Mnull \oplus \C^K, \qquad
  q_0(u):= \sum_j \normsqr[\Ij] {u_j'}\,,
\end{equation}
% ------------- %
where the form domain $\mathcal D_0$ of $q_0$ is given by those
functions $u=((u_j)_{j\in J},(u_k)_{k\in K})$ such that
% ------------- %
\begin{equation}
  \label{def:lim.dom.border}
  u \in \Sob \Mnull \oplus \C^K \qquad \text{and} \qquad
  (\vol \mVk)^{1/2} u_j(v_k) = u_k
\end{equation}
% ------------- %
for all $j \in J_k$ and $k \in K$, i.e., values of the functions at
the edge endpoints $v_k\equiv x^0_{jk}$ are now coupled with the
additional wave function components; recall that $\mVk$ denotes
the manifold $\mVepsk$ with $\eps=1$. The corresponding operator
$Q_0$ is given by
% ------------- %
\begin{equation}
  \label{def:lim.op.border}
  Q_0 u =
  \biggl(
    \Bigl(
       -\frac1{p_j} (p_j u_j')'
    \Bigr)_{\!j},
    \Bigl(
       - (\vol \mVk)^{-\frac12} \sum_{j \in J_k} p_j(v_k)  u_j'(v_k)
    \Bigr)_{\!k}
  \biggr) \,;
\end{equation}
% ------------- %
it depends parametrically on $\vol(\mVk)$ but we refrain from
marking this fact explicitly. Again, this operator has a purely
discrete spectrum provided the graph $\Mnull$ is finite.

As we have said, $Q_0$ is not a graph operator with the conditions
\eqref{eq:delta}. Nevertheless, there is a similarity between the
two noticed by Kuchment and Zeng in \cite{kuchment-zeng:03}. To
solve the spectral problem $Q_0 u=\lambda u$ one has to find
$(u_j)_{j\in J}$ such that $-(p_j u_j')'/p_j = \lambda u_j$ and at
the vertices the functions satisfy the conditions
% ------------- %
\begin{equation}
  \label{def:delta.spectral}
  \sum_{j \in J_k} p_j(v_k) u_j'(v_k) = - \lambda (\vol \mVk) u(v_k)\,.
\end{equation}
% ------------- %
This looks like \eqref{eq:delta}, the difference is that the
coefficient at the right-hand side is not a constant but a
multiple of the spectral parameter; in physical terms one may say
that the coupling strength at a vertex is proportional to the
energy.

After this digression let us return to the limiting properties. We adopt again
the assumption~\eqref{eq:est.a.r} in this section. Instead of
(\ref{eq:met.vertex.slow}) we suppose now that on the vertex neighbourhood the
metric satisfies the relation
% ------------- %
\begin{equation}
  \label{eq:met.vertex.border}
  g_\eps = \eps^{2\alpha} g + o(\eps^{2\alpha})
  \qquad \text{on $\mVk$}
\end{equation}
% ------------- %
with
% ------------- %
 \begin{equation}
  \label{eq:est.alpha.border}
  \alpha = \frac {d-1} d\,,
\end{equation}
which corresponds to the above mentioned equal decay rate for the
volume of the edge and vertex neighbourhoods. In particular, we have
\begin{equation}
  \label{eq:norm.quad.border}
  \normsqr[\mVeps] u = \eps^{\alpha d} (1+o(1)) \normsqr[\mV] u
  \quad \text{and} \quad
  \normsqr[\mVeps] {du} = \eps^{\alpha (d-2)} (1+o(1)) \normsqr[\mV] {du}
\end{equation}
and
\begin{equation}
  \label{eq:vol.border}
  \vol (\mVeps) = \eps^{\alpha d} (1+o(1)) \vol (\mV)
\end{equation}
for each $\mV=\mVk$ as in Lemmas~\ref{lem:metric} and~\ref{lem:est.norm.quad}.

%------------------------------------------------------------
\subsection{Convergence of the spectra}
%------------------------------------------------------------

With the above prerequisites we can finally formulate the main
result of this section:
% ------------- %
\begin{thm}
  \label{thm:ev.conv.border}
  Under the stated assumptions $\EW k \Meps \to \EW k {Q_0}$ as $\eps\to 0$.
\end{thm}
% ------------- %

\noindent To prove it, our aim is again to find a two sided
estimate on each eigenvalue $\EW k \Meps$ by means of $\EW k
{Q_0}$ with an error which is $o(1)$ w.r.t.\ the parameter $\eps$.

%------------------------------------------------------------
\subsection{An upper bound}
%------------------------------------------------------------

Again, we first show the easier upper eigenvalue estimate:
% ------------- %
\begin{thm}
\label{thm:ev.above.border}
  $\EW k \Meps \le \EW k {Q_0} + o(1)$ holds as $\eps\to 0$.
\end{thm}
% ------------- %
\noindent We define the transition operator by
% ------------- %
\begin{equation}
\label{eq:trans.op.above.border}
  \Phi_\eps u (z):=
  \begin{cases}
    \vol(\mVepsk)^{-1/2} u_k & \text{if $z \in \Vk$},\\
    \begin{aligned}
      \eps^{-m/2} & u_j(x)  +
       \rho(x) \times {} \\ & {} \times
       \bigl(
          \vol(\mVepsk)^{-1/2} u_k - \eps^{-m/2} u_j(x^0)
       \bigr)
    \end{aligned} &
           \text{if $z=(x,y) \in \Uj$}
  \end{cases}
\end{equation}
% ------------- %
for any $u \in \mathcal D_0$, where $\rho$ is a smooth function as
in~\eqref{eq:smooth.fct} and $x^0=x^0_{jk}$ denotes the endpoint of
the half-edge $\Ijk$ away from the vertex $v_k$.
Theorem~\ref{thm:ev.above.border} is then implied by
Lemma~\ref{lem:main} in combination with the following result.

% ------------- %
\begin{lem}
  \label{lem:above.border}
  We have $\Phi_\eps u \in \Sob \Meps$, i.e., $\Phi_\eps$ maps the quadratic
  form domain $\mathcal D_0$ into the quadratic form domain of the Laplacian
  on the manifold. Furthermore,
% ------------- %
  \begin{align}
    \normsqr[\HS_0] u - \normsqr[\Meps] {\Phi_\eps u} &\le
    o(1) \, \normsqr[\HS_0] u\\
    \normsqr[\Meps] {d \, \Phi_\eps u} - q_0(u) & \le
    o(1) \, (\normsqr[\HS_0] u + q_0(u))
  \end{align}
% ------------- %
\end{lem}
% ------------- %
\begin{proof}
  The argument is analogous to the proof of Lemma~\ref{lem:above.slow}. The
  only difference is that we need the following estimate
% ------------- %
  \begin{displaymath}
    \eps^m \bigl|
       \vol(\mVepsk)^{-1/2} u_k - \eps^{-m/2} u_j(x^0)
    \bigr|^2 =
    \bigl| \eps^{m/2} (\vol \mVepsk)^{-1/2} - (\vol \mVk)^{-1/2} \bigr|^2 \,
      | u_k |^2 %\le o(1) \normsqr[\HS_0] u
  \end{displaymath}
% ------------- %
  since $u \in \mathcal D_0$. The last difference is of order $o(1)$
  by~\eqref{eq:vol.border}.
\end{proof}
% ------------- %

%------------------------------------------------------------
\subsection{A lower bound}
%------------------------------------------------------------

The estimate on $\EW k \Meps$ from below can be found in analogy
with the slowly decaying case in Section~\ref{sec:slow.decay}.
Furthermore, we need the following averaging operator
% ------------- %
\begin{displaymath}
  C_k^- u := \frac 1 {\vol \mVk} \int_\mVk u \, d\mVk.
\end{displaymath}
% ------------- %
Since we have an exact scaling of the metric of order $\eps^\alpha$
by~\eqref{eq:est.alpha.border}, we also could use the $\eps$-depending
manifold $\mVepsk$ here (cf.\ also Remark~\ref{rem:no.eps}).

% ------------- %
\begin{lem}
  \label{lem:diff.av.border}
  For all $u \in \Sob \mVepsk$ we have
% ------------- %
  \begin{displaymath}
    \bigl| C^-_k u - N_j u(x^0) \bigr|^2 \le
    o(\eps^{-m}) \, \normsqr[\Vepsk] {du}
  \end{displaymath}
% ------------- %
\end{lem}
% ------------- %
\begin{proof}
  We have
% ------------- %
  \begin{displaymath}
    \bigl| C^-_k u - N_j u(x^0) \bigr| \le
    \bigl| C^-_k u - N_j u(x^+) \bigr| +
    \bigl| N_j u(x^+) - N_j u(x^0) \bigr|.
  \end{displaymath}
% ------------- %
  The first difference can be estimated in the same way as
  Lemma~\ref{lem:diff.av} (replacing $\Vk$ by $\mVk$ and using
  estimate~\eqref{eq:norm.quad.border}, i.e., we arrive at
% ------------- %
  \begin{displaymath}
     \bigl| C^-_k u - N_j u(x^+) \bigr|^2 \le
     O(\eps^{-(d-2)\alpha}) \, \normsqr[\mVeps]{du};
  \end{displaymath}
% ------------- %
  recall that now we have $\alpha d = m$.  For the second difference,
  use~\eqref{eq:diff.av.slow}.
\end{proof}
% ------------- %

Similarly to Lemma~\ref{lem:diff.vol.av.slow} we can prove: 
% ------------- %
\begin{lem}
  \label{lem:diff.vol.av.border}
  For all $u \in \Sob \mVeps$, we have
  \begin{displaymath}
    \normsqr[\mVeps] u - \normsqr[\mVeps]{C^- u} \le
    O(\eps^\alpha) (\normsqr[\mVeps] u + \normsqr[\mVeps] {du}).
  \end{displaymath}
\end{lem}
% ------------- %
% \begin{proof}
%   The proof is similar to the proof of Lemma~\ref{lem:diff.vol.av.slow}.
% %   The proof is similar to the proof of Lemma~\ref{lem:diff.vol.av.slow}:
% %   inequality~\eqref{ineq:norm} together with~\eqref{eq:norm.quad.border}
% %   and 
% %   Lemma~\ref{lem:minmax.2nd.neu} for $X=\mV$ implies
% % % ------------- %
% %   \begin{multline*}
% %     \normsqr[\mVeps] u - \normsqr[\mVeps] {C^- u} \le 
% %       \frac 1 \delta
% %            \eps^{\alpha d} (1+o(1)) 
% %            \normsqr[\mV] {u - C^-u} + \delta ( \normsqr[\mVeps] u
% %            + \vol \mVeps |C^- u|^2) \\ \le 
% %       \frac {\eps^{2\alpha}(1+o(1))} {\delta \EWN 2 \mV}
% %       \normsqr[\mVeps] {du} + 
% %       \delta ( \normsqr[\mVeps] u + \vol \mVeps \normsqr[\mV] u)
% %   \end{multline*}
% % % ------------- %
% %   Applying~\eqref{eq:vol.border} and \eqref{eq:norm.quad.border} once more,
% %   the result follows setting $\delta=\eps^\alpha$.
% \end{proof}

Now we define the transition operator by
% ------------- %
\begin{equation}
  \label{eq:trans.op.below.border}
  \begin{split}
    (\Psi_\eps u)_j (x) &:=
      \eps^{m/2} \Bigl(N_j u (x) +
            \rho(x) \bigr(C^-_k u - N_j u (x^0) \bigr)\Bigr)
      \quad \text{for $x \in I_{jk}$} \\
    (\Psi_\eps u)_k &:=
      \eps^{m/2} (\vol \mVk)^{1/2} C^-_k u
  \end{split}
\end{equation}
% ------------- %
where $\rho$ is a smooth function as in~\eqref{eq:smooth.fct} and
$x^0=x^0_{jk}$ denotes the endpoint of the half-edge
$\Ijk$ corresponding to the vertex $v_k$.
% ------------- %
\begin{lem}
  \label{lem:below.border}
  We have $\Psi_\eps u \in \mathcal D_0$ if $u \in \Sob \Meps$.
  Furthermore,
  % ------------- %
  \begin{align}
    \normsqr[\Meps] u - \normsqr[\HS_0] {\Psi_\eps u} &\le
    o(1) \bigl(
       \normsqr[\Meps] u + \normsqr[\Meps] {du}
    \bigr) \\
    q_0(\Psi_\eps u) - \normsqr[\Meps] {du} &\le
    o(1) \bigl(
       \normsqr[\Meps] u + \normsqr[\Meps] {du}
    \bigr)
  \end{align}
  % ------------- %
  for all $u \in \Sob \Meps$.
\end{lem}
% ------------- %
\begin{proof}
  The arguments are the same as in the proof of
  Lemma~\ref{lem:below.slow}.
  For the vertex contribution, we need the estimate
% ------------- %
  \begin{displaymath}
    \normsqr[\mVepsk] u - \eps^m (\vol \mVk) |C^-_k u|^2  =
    (\normsqr[\mVepsk] u - \normsqr[\mVepsk] {C^-_k u}) +
    \Bigl(  \frac{\vol \mVepsk}{\eps^m \vol \mVk} - 1 \Bigr)
      \eps^m \normsqr[\mVk] {C^-_k u}.
  \end{displaymath}
% ------------- %
  The first difference can be treated with Lemma~\ref{lem:diff.vol.av.border}
  and leads to an error term $O(\eps^\alpha)$. The second term is of order
  $o(1) \normsqr[\mVepsk] u$ by~\eqref{eq:norm.quad.border},
  \eqref{eq:vol.border} and Cauchy-Schwarz.  Furthermore,
  Corollary~\ref{cor:rest.small} is also true in this setting (independent on
  the particular $\alpha$). We also need Lemma~\ref{lem:diff.av.border}.
\end{proof}

Using Lemma~\ref{lem:below.border} we arrive at the sought lower bound.
Again, the error term $\eta_k$ in~\eqref{eq:eta.k} can be
estimated by some $\eps$-independent quantity because
$\lambda_k=\EW k \Meps \le c_k$ by Theorem~\ref{thm:ev.above.border}.
% ------------- %
\begin{thm}
\label{thm:ev.below.border}
  We have $\EW k {Q_0} \le \EW k \Meps + o(1)$.
\end{thm}
% ------------- %
\noindent Theorem~\ref{thm:ev.conv.border} now follows easily by
combining the last result with Theorem~\ref{thm:ev.above.border}.

%------------------------------------------------------------
\section{Non-decaying vertex volume}
\label{sec:alpha.null}
%------------------------------------------------------------

In this section, we treat the case when the vertex volume does not tend to
$0$. In some sense, this case corresponds to $\alpha = 0$ in the previous
notation but we need more assumptions to precise the convergence of the
manifold $\Vepsk$ to a manifold $\Vnullk$ as $\eps \to 0$. We cite only the
result here since it has already been presented in~\cite{post:03a} or with a
more detailed proof in~\cite{post:00}. A related result corresponding to the
embedded case (see Example~\ref{ex:embedded}) as in \cite{kuchment-zeng:01}
was proven by Jimbo and Morita in \cite{jimbo-morita:92} or for manifolds
(with non-smooth junctions between edge and vertex neighbourhoods) by Ann\'e
and Colbois in \cite{anne-colbois:95}.

Furthermore, we assume that the transversal direction is a sphere, i.e.,
$F=\Sphere^m$. Let $\Vnullk$ be a compact $d$-dimensional manifold without
boundary for $k \in K$. To each edge $j \in J_k$ emanating from the vertex
$v_k$, we associate a point $x_{jk}^0 \in \Vnullk$ such that $x_{jk}^0$ ($j
\in J_k$) are mutually distinct points with lower bound $2 \eps_0>0$ on their
distance to each other. We assume for simplicity that the metric at
$x^0=x_{jk}^0$ is locally flat within a distance $\eps_0$ from $x^0$ (the
general case can be found in~\cite{post:03a}). Then the metric in polar
coordinates $(x,y) \in (0,\eps_0) \times \Sphere^m$ looks locally like
\begin{displaymath}
  g = dx^2 + x^2 \, h_y
\end{displaymath}
%(cf.~\cite[Prop.~E.III.7]{bgm:71})
where $h_y$ is the standard metric on $\Sphere^m$. Modifying the factor before
$h_y$, we define a new metric by
\begin{displaymath}
  g_\eps = dx^2 + r_\eps^2(x) \, h_y
\end{displaymath}
with a smooth monotone function $\map {r_\eps} {(0,\eps_0)}{(0,\infty)}$ such
that
\begin{displaymath}
  r_\eps(x) = 
  \begin{cases}
    \eps & \text{for $0 < x < \eps/2$}\\
    x    & \text{for $2\eps < x < \eps_0$}.
  \end{cases}
\end{displaymath}
We denote the (completion of the) manifold $(\Vnullk \setminus \bigcup_{j \in
  J_k} \set{x_{jk}^0}{ j \in J_k}, g_\eps)$ by $\Vepsk$. Note that this
manifold has $|J_k|$ attached cylindrical ends of order $\eps$ at each point
$x_{jk}^0$. Now we can construct the graph-like manifold $\Meps$ as in
Section~\ref{sec:graph.mfd}.

As in the slowly decaying case of Section~\ref{sec:slow.decay} the limit
operator
\begin{displaymath}
  Q_0 := \bigoplus_{j \in J} \laplacianD \Ij \oplus
         \bigoplus_{k \in K} \laplacian \Vnullk
\end{displaymath}
decouples and the next result follows (cf.~\cite[Theorem~1.2]{post:03a}
or~\cite{post:00}):
\begin{thm}
  \label{thm:ev.conv.null}
  We have $\EW k \Meps \to \EW k {Q_0}$ as $\eps \to 0$.
\end{thm}

%------------------------------------------------------------
\section{Applications}
\label{sec:applications}
%------------------------------------------------------------

Finally we comment on consequences of the spectral convergence. We begin with
a general remark stating that we only have uniform control over a
\emph{compact} spectral interval:
\begin{rem}
  \label{rem:not.uniform}
  Note that the convergence $\EW k \Meps \to \EW k \Mnull$ cannot be uniform
  in $k \in \N$: if this were the case, the theta-function
  \begin{displaymath}
    \Theta_\eps(t):= \tr \eu^{-t\laplacian \Meps} = \sum_k \eu^{-t \EW k \Meps}
  \end{displaymath}
  would converge to $\Theta_0(t)$. But Weyl asymptotics are different in the
  two cases, 
  \begin{displaymath}
    \Theta_\eps(t) \sim \frac{\vol_d \Meps}{(4\pi t)^{d/2}}, 
      \qquad \text{whereas} \qquad
    \Theta_0(t) \sim \frac{\vol_1 \Mnull}{(4 \pi t)^{1/2}}
  \end{displaymath}
  as $t \to 0$ (cf.~\cite[Sec.~VI.4]{chavel:84} and~\cite[Thm.~1]{roth:84}).
  Recall that $d \ge 2$ and $\vol_1 \Mnull := \sum_j \ell_j$, i.e. the sum
  over the length of each edge.
\end{rem}

%------------------------------------------------------------
\subsection{Periodic graphs}
\label{ssec:per.graph}
%------------------------------------------------------------
Suppose we have an infinite graph $X_0$ on which a discrete, finitely
generated group $\Gamma$ operates such that the quotient $M_0 := X_0 / \Gamma$
is a finite graph. In the same way as in the previous sections, we can
associate a family of graph-like compact manifolds $\Meps$ to the graph $M_0$.
By a lifting procedure we obtain a (non-compact) covering manifold $X_\eps$ of
$\Meps$ with deck transformation group $\Gamma$, i.e., $\Meps$ is isometric to
$X_\eps / \Gamma$. Furthermore, $X_\eps$ is a graph-like manifold collapsing
to the infinite graph $X_0$.

We are interested in spectral properties of the non-compact manifolds
$X_\eps$.  Assuming that $\Gamma$ is abelian, we can apply Floquet theory (for
a non-commutative version see~\cite{lledo-post:pre04}).  Instead of
investigating $\laplacian {X_\eps}$ we analyze a family of operators
$\laplacianT \Meps$, $\theta \in \hat \Gamma$, where $\hat \Gamma$ is the dual
group, i.e., the group of homomorphisms from $\Gamma$ into the unit circle
$\Torus^1$. The operator $\laplacianT \Meps$ acts on a complex line bundle
over the compact manifold $\Meps$, or equivalently, over the closure of a
fundamental domain $D_\eps \subset X_\eps$ with $\theta$-periodic boundary
conditions. We call the closure $\overline D_\eps$ a \emph{period cell} and
denote it also by $\Meps$ (for details see e.g.\ \cite{reed-simon-4}
or~\cite{post:03a}). The direct integral decomposition implies
\begin{displaymath}
  \spec \laplacian {X_\eps} = \bigcup_{k \in \N} B_k(\eps), \qquad
  B_k(\eps) := \set{\EWT k \Meps} {\theta  \in \hat \Gamma}
\end{displaymath}
where $B_k(\eps)$ is a compact subset of $[0,\infty)$, called the \emph{$k$-th
  band}.\footnote{Note that $\hat \Gamma$ is connected iff $\Gamma$ is torsion
  free, e.g., if $\Gamma=\Z \times \Z_2$ then $\hat \Gamma \cong \Torus^1
  \times \Z_2$ which is homeomorphic to two disjoint copies of the unit circle
  $\Torus^1$. Therefore, the bands $B_k(\eps)$ being the continuous image of
  $\hat \Gamma$ under the map $\theta \mapsto \EWT k \Meps$ need not to be
  intervals. } A similar assertion holds for the limit operator on $X_\eps$.

%------------------------------------------------------------
\subsection{Spectral gaps}
%------------------------------------------------------------
We are interested in the existence of \emph{spectral gaps} of the operator
$\laplacian {X_\eps}$, i.e., the existence of an interval $[a,b]$, $0<a<b$,
such that $\spec \laplacian {X_0} \cap [a,b] = \emptyset$. Note that $\spec
\laplacian {X_\eps}$ is purely essential.
\begin{thm}
  \label{thm:gaps}
  We have $\EWT k \Meps \to \EWT k {Q_0}$ for $\eps \to 0$ \emph{uniformly} in
  $\theta \in \hat \Gamma$. Furthermore,
  \begin{displaymath}
    B_k(\eps) \cap B_{k+1}(\eps) = \emptyset 
    \qquad \text{if} \qquad
    B_k(0) \cap B_{k+1}(0) = \emptyset
  \end{displaymath}
  provided $\eps$ is small enough. In particular, an arbitrary (but finite)
  number of gaps open up in the spectrum of $\laplacian {X_\eps}$ provided the
  limit operator $Q_0$ has enough gaps and $\eps$ is small enough.
\end{thm}
\begin{proof}
  The spectral convergence can be proven in the same way as in the previous
  sections. Note that the error terms converge \emph{uniformly} in $\theta \in
  \hat \Gamma$ since all error bounds are independent of $\theta$.  The only
  point where $\theta$ enters is the error estimate~\eqref{eq:eta.k} for the
  lower eigenvalue estimate.  In this case, we argue as follows: we have $\EWT
  k \Meps \le \EWD k \Meps$, i.e., the Dirichlet Laplacian eigenvalues form an
  upper bound on the $\theta$-periodic eigenvalues. Here, we pose Dirichlet
  boundary conditions on the boundary of the period cell. Furthermore, $\EWD k
  \Meps \to \EWD k \Mnull$ by the same arguments as in the previous sections.
  Therefore, we can choose $\lambda_k = \EWT k \Meps \le \EWD k \Meps \le 2
  \EWD k \Mnull$ in~\eqref{eq:eta.k} \emph{independently} of $\theta$.
\end{proof}
Note that we cannot expect to show the existence of \emph{infinitely many}
gaps in $X_\eps$ even if $\spec \laplacian {X_0}$ has infinitely many gaps
since the convergence is not uniform in $k$ (cf.\ 
Remark~\ref{rem:not.uniform}). This is related to the deep open problem about
the validity of Bohr-Sommerfeld conjecture on such periodic manifolds.
\begin{rem}
  \label{rem:overlap}
  If two neighboured bands $B_k(0)$ and $B_{k+1}(0)$ \emph{overlap}, i.e.,
  intersect in a set of positive length, the same is true for $B_k(\eps)$ and
  $B_{k+1}(\eps)$ provided $\eps$ is small enough. In contrast, if the bands
  intersect only in \emph{one} point, i.e., if they \emph{touch} each other,
  we cannot say anything about the (non-)existence of gaps in the spectrum of
  $\laplacian {X_\eps}$.
\end{rem}

For the rest of this section we discuss examples for which
Theorem~\ref{thm:gaps} applies.

%------------------------------------------------------------
\subsection{Decoupling limit operators}
%------------------------------------------------------------
Suppose that our graph-like periodic manifold $X_\eps$ is constructed as in
Section~\ref{sec:slow.decay} or~\ref{sec:alpha.null}. In this case, the limit
operator is a direct sum of the limit operator on the quotient $\Mnull$ since
the limit operator \emph{decouples}. Therefore, the bands $B_k(0)$ degenerate
to the points $\EW k {Q_0}$ where $Q_0$ is given as in
Sections~\ref{sec:slow.decay} or~\ref{sec:alpha.null} and the limit operator
on $X_0$ has infinitely many gaps. This means, in particular, that the limit
spectrum is not absolutely continuous, while those of the approximating
operators may be. Furthermore, Theorem~\ref{thm:gaps} applies in this case.

%------------------------------------------------------------
\subsection{Cayley graphs and Kirchhoff boundary conditions}
%------------------------------------------------------------
In the following three subsections, we give examples of graph-like manifolds
with fast decaying vertex volume as constructed in
Section~\ref{sec:fast.decay} such that $\laplacian {X_\eps}$ has spectral
gaps. In this case, the limit operator is the Laplacian $\laplacian {X_0}$ on
the graph $X_0$ with Kirchhoff boundary conditions as in~\eqref{eq:kirchhoff}.
We want to calculate the spectrum of $\laplacian {X_0}$ for certain graphs
$X_0$. For simplicity, we assume that $p_j \equiv 1$ and that each edge has
length $1$.

Suppose that $\Gamma$ is an abelian, finitely generated discrete group.
Therefore,
\begin{displaymath}
  \Gamma \cong \Z^{r_0} \times \Z_{p_1}^{r_1} \times \dots 
            \times \Z_{p_a}^{r_a}
\end{displaymath}
where $\Z_p$ is the cyclic group of order $p$.  Furthermore, $r_0>0$ since
$X_0$ is non-compact and $X_0/\Gamma$ is compact. Denote $r:=r_0 + r_1 + \dots
+ r_a$.

\begin{figure}[h]
  \begin{center}
%------------------------------------------------------------
%        \input{cayley.pstex_t}
\begin{picture}(0,0)%
  \includegraphics{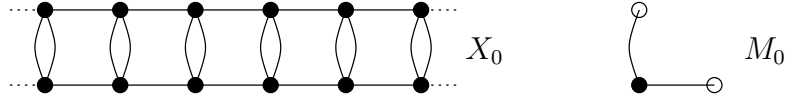}%
\end{picture}%
\setlength{\unitlength}{4144sp}%
\begin{picture}(4422,728)(214,-286)
  \put(2971, 100){$X_0$}%
  \put(4636, 100){$M_0$}%
\end{picture}
%------------------------------------------------------------
    \caption{The Cayley graph associated to the group $\Gamma = \Z \times
      \Z_2$ and the corresponding period cell. Note that $\laplacian {X_0}$
      has no spectral gaps}
    \label{fig:cayley}
  \end{center}
\end{figure}

We assume that $X_0$ is the (metric) Cayley graph associated to $\Gamma$
w.r.t.\ the canonical generators $\eps_1, \dots, \eps_r$ ($\eps_j$ equals $1$
at the $j$-th component and $0$ otherwise), i.e., the set of vertices is
$\Gamma$ and two vertices $\gamma_1,\gamma_2$ are connected iff
$\gamma_2=\eps_j \gamma_1$ for some $1 \le j \le r$ (see
Figure~\ref{fig:cayley}). Note that $X_0$ is $2r$-regular, i.e., each vertex
meets $2r$ edges.
We want to calculate the eigenfunctions and eigenvalues of the
$\theta$-periodic operator $\laplacianT \Mnull$, i.e., functions $u_j$ on $\Ij
\cong [0,1]$ satisfying $-u_j''=\lambda u_j$ with the boundary conditions
\begin{equation}
  \label{eq:bd.cond}
  u_j(0) = u(0), \quad
  \eu^{-\im\theta_j} u_j(1) = u(0) \quad \text{and} \quad
  \sum_{k=1}^r \bigl( \eu^{-\im \theta_k} u_k'(1) - u_k'(0) \bigr) = 0
\end{equation}
for all $j=1, \dots, r$. Here, $\theta \in \Torus^{r_0} \times
\Torus_{p_1}^{r_1} \times \dots \times \Torus_{p_a}^{r_a}$ where $\Torus_p :=
\set{\xi \in \R/\Z}{\eu^{\im \xi p} = 1}$ is the group of $p$-th unit roots
(isomorphic to $\Z_p$). Note that we have identified $\theta \in \Torus^r$
with $\gamma \mapsto \eu^{\im \theta \cdot \gamma} \in \hat \Gamma$.

If $\lambda=\omega^2 > 0$ (and $\omega>0$) we make the Ansatz
\begin{equation}
  \label{eq:ansatz}
  u_j(x):=Z \cos(\omega x) + A_j \sin(\omega x).
\end{equation}
Non-trivial solutions of the eigenvalue problem exist
iff
% \begin{displaymath}
%   2 (\sin^{r-1}\omega) \eu^{-\im \sum_k \theta_k}
%      \sum_k (\cos \omega - \cos \theta_k) = 0,
% \end{displaymath}
% i.e.\ if
$\omega = \ell \pi$, $\ell \in \N$, or
% \begin{displaymath}
%   u_j(x):=Z \cos(\omega x) + A_j \sin(\omega x)
% \end{displaymath}
% and arrive at the coefficient matrix $M(\omega)$ given by
% \begin{displaymath}
% {\scriptsize 
%   \begin{pmatrix}
%     \eu^{-\im \theta_1} \sin \omega & 0 & \cdots & 0 & 
%                                        (\eu^{-\im \theta_1} \cos \omega - 1) \\
%     0     & \eu^{-\im \theta_2} \sin \omega  & \cdots & 0 & 
%                                        (\eu^{-\im \theta_2} \cos \omega - 1) \\
%     \vdots &&& 0 & \vdots \\
%     0 & 0 & \cdots & \eu^{-\im \theta_r} \sin \omega &  
%                                        (\eu^{-\im \theta_r} \cos \omega - 1) \\
%      (\eu^{-\im \theta_1} \cos \omega - 1) &  
%                                        (\eu^{-\im \theta_2} \cos \omega - 1) &
%      \cdots & (\eu^{-\im \theta_r} \cos \omega - 1) &
%       (- \sin \omega \sum_{k=1}^r \eu^{-\im \theta_k})
%   \end{pmatrix}
% }
% \end{displaymath}
% for the variables $A_1, \dots, A_r, Z$.  A direct calculation shows that
% \begin{displaymath}
%   \det M(\omega) = 2 (\sin^{r-1}\omega) \eu^{-\im \sum_k \theta_k}
%      \sum_k (\cos \omega - \cos \theta_k).
% \end{displaymath}
% Non-trivial solutions of the eigenvalue problem exist iff $\det M(\omega)=0$,
% i.e. if $\omega = \ell \pi$, $\ell \in \N$, or
\begin{equation}
  \label{eq:omega.theta}
  \cos \omega = \frac 1 r \sum_{k=1}^r \cos \theta_k.
\end{equation}
The solutions $\omega = \ell \pi$ correspond to Dirichlet eigenfunctions on
each edge and produce therefore bands degenerated to a point $\{ (\ell\pi)^2
\}$. The multiplicity is $r-1$ provided $\theta \ne 0$ (if $\ell$ is even)
resp.\ $\theta \ne \pi$ (if $\ell$ is odd) and $r+1$ if $\theta = 0$ resp.\ 
$\theta = \pi$ (modulo $2\pi$).  If $\omega \ne \ell \pi$, the eigenvalues are
simple. Note that the bands at $\omega^2 = (\ell \pi)^2$ do not overlap, but
\emph{touch} each other.

For $\omega=0$, we need a special Ansatz.  The only possibility is the case of
periodic boundary conditions ($\theta=0$); the eigenvalue is simple.

\begin{thm}
  \label{thm:gaps.ex}
  If one of the orders $p_1, \dots, p_a$ is odd, the operator $\laplacian
  {X_0}$ has infinitely many spectral gaps below and above $(2\ell+1)^2\pi^2$
  ($\ell=0,1,\dots$). In particular, Theorem~\ref{thm:gaps} applies.
  Furthermore, the bands $\{(2\ell+1)^2\pi^2\}$ are degenerated to a point and
  have multiplicity $r-1$. The gap length increases as $\ell \to \infty$.
  
  If all orders $p_1, \dots, p_a$ are even then $\spec \laplacian {X_0} = [0,
  \infty)$.
\end{thm}
\begin{proof}
  We analyze the behaviour of $\omega$ in dependence of the continuous
  parameters $\theta_1, \dots, \theta_{r_0} \in \Torus^{r_0}$ given by the
  relation~\eqref{eq:omega.theta}. We have gaps iff $\frac 1 r \sum_{k=1}^r
  \cos \theta_k$ in~\eqref{eq:omega.theta} does not cover the whole interval
  $[-1,1]$. We reach the maximal value $1$ iff all $\theta_j=0$ ($j=1, \dots,
  r$) and the minimal value $-1$ iff all $\theta_j=\pi$ ($j=1, \dots, r$). The
  latter can only occur if all group orders are even. Note that in this case,
  the whole interval $[-1,1]$ can be covered by an appropriate choice of the
  $\theta_j$'s, $j=r_0+1, \dots, r$. If one $p_j$ is odd, there exists $\eps >
  0$ such that \eqref{eq:omega.theta} has no solution provided $(2\ell+1)\pi -
  \eps < \omega < (2\ell+1)\pi + \eps$.
\end{proof}
We cannot say anything about the (non-)existence of gaps in the case when all
orders $p_1, \dots, p_a$ are even. If e.g.\ $\Gamma = \Z \times \Z_2$, the
bands do not overlap, but touch each other and fill the whole half line
$[0,\infty)$ (cf.\ Remark~\ref{rem:overlap}).

%------------------------------------------------------------
\subsection{Non-commutative groups}
%------------------------------------------------------------
We comment briefly on a similar result for certain non-commutative groups
$\Gamma$. Here, $\hat \Gamma$ consists of (equivalence classes of) irreducible
unitary representations (cf.~\cite{lledo-post:pre04}). A simple example is
given by $\Gamma = \Z \times D_n$, where $D_n$ denotes the dihedral group of
order $2n$ generated by $\alpha, \beta$ with $\alpha^2=e$, $\beta^n=e$ and
$\alpha \beta = \beta^{-1} \alpha$.
\begin{figure}[h]
  \begin{center}
%------------------------------------------------------------
% \newcommand{\color}[2][{}]{}
%         \input{graph5a.pstex_t}
\begin{picture}(0,0)%
  \includegraphics{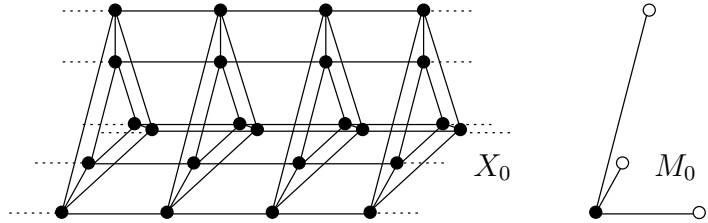}%
\end{picture}%
\setlength{\unitlength}{4144sp}%
\begin{picture}(4197,1298)(157,-562)
  \put(2971,-286){$X_0$}%
  \put(4051,-286){$M_0$}%
\end{picture}
%------------------------------------------------------------
    \caption{The Cayley graph associated to the group $\Gamma=\Z \times D_3$
      where $D_3$ is the dihedral group of order $6$. The corresponding
      Laplacian has spectral gaps.}
    \label{fig:nc.cayley}
  \end{center}
\end{figure}
In this case the Laplacian on the corresponding Cayley graph (cf.\ 
Figure~\ref{fig:nc.cayley}) has infinitely many spectral gaps below and above
$(2\ell+1)^2\pi^2$, $\ell=0,1,\dots,$ if $n$ is odd. If $n$ is even, $\spec
\laplacian {X_0} = [0,\infty)$. For a related family of sleeve manifolds in
case of odd $n$ there is an arbitrary large number of open gaps provided the
sleeve radius is small enough.

%------------------------------------------------------------
\subsection{Cayley graphs with loops}
\begin{figure}[h]
  \begin{center}
%------------------------------------------------------------
%        \input{cayley.loop.pstex_t}
\begin{picture}(0,0)%
\includegraphics{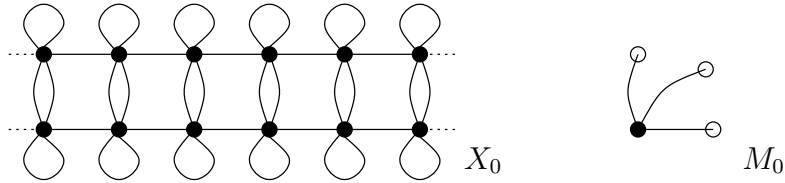}%
\end{picture}%
\setlength{\unitlength}{4144sp}%
\begin{picture}(4422,1074)(214,-373)
  \put(2971,-286){$X_0$}%
  \put(4636,-286){$M_0$}%
\end{picture}
%------------------------------------------------------------
    \caption{The Cayley graph associated to the group $\Gamma = \Z \times
      \Z_2 \times \Z_1$, where the trivial group $\Z_1$ leads to the
      attachment of a loop at each vertex. On the right, the corresponding
      period cell is shown. Note that $\laplacian {X_0}$ has spectral gaps in
      contrast to the example without loops.}
    \label{fig:cayley.loop}
  \end{center}
\end{figure}

%------------------------------------------------------------
If we set one (or more) of the group orders $p_j$ equal to $1$ we formally
attach a loop (or more) at each vertex (see Figure~\ref{fig:cayley.loop}).
\begin{thm}
  \label{thm:gaps.ex.loop}
  The Laplacian of a Cayley graph associated to an arbitrary finitely
  generated abelian discrete group $\Gamma$ has infinitely many spectral gaps
  provided we attach at each vertex a fixed number of loops.
\end{thm}
\begin{proof}
  Formally, the assertion follows from Theorem~\ref{thm:gaps.ex}. Note that
  $\hat \Z_1 = \{0\}$, i.e., the corresponding component of $\theta$ cannot be
  $\pi$ and therefore, the minimum $-1$ cannot be achieved
  in~\eqref{eq:omega.theta}.
\end{proof}
This is an analogue of gap generation by decoration of the graph as discussed
by Aizenman and Schenker in \cite{aizenman-schenker:00}.

%------------------------------------------------------------
\subsection{Cayley graphs and the borderline case}
%------------------------------------------------------------
In the borderline case, the eigenvalue problem of the limit operator is more
complicated. Here, functions $u_j$ on $\Ij
\cong [0,1]$ satisfy $-u_j''=\lambda u_j$ with the boundary conditions
\begin{equation}
  \label{eq:bd.cond.2}
  u_j(0) = u(0), \quad
  \eu^{-\im\theta_j} u_j(1) = u(0) \quad \text{and} \quad
   \sum_{k=1}^r \bigl( \eu^{-\im \theta_k} u_k'(1) - u_k'(0) \bigr) = c \lambda
  u(0)
\end{equation}
for all $j=1, \dots, r$ where $c^2$ is the volume of the (unscaled) vertex
neighbourhood (cf.~\eqref{def:lim.op.border}). Again, with the
Ansatz~\eqref{eq:ansatz}, non-trivial solutions of the eigenvalue problem
exist iff
% \begin{displaymath}
%   (\sin^{r-1}\omega) \eu^{-\im\sum_k \theta_k}
%      \bigl( 
%         2 \sum_k (\cos \omega - \cos \theta_k) - c \omega \sin \omega 
%      \bigr) = 0,
% \end{displaymath}
% i.e.\ if
$\omega = \ell \pi$, $\ell \in \N$, or
% \begin{displaymath}
%   u_j(x):=Z \cos(\omega x) + A_j \sin(\omega x)
% \end{displaymath}
% with $\lambda=\omega^2 > 0$ and arrive at the coefficient matrix $M_c(\omega)$
% given by
% \begin{displaymath}
% {\scriptsize  \begin{pmatrix}
%     \eu^{-\im \theta_1} \sin \omega & 0 & \cdots & 0 & 
%                                        (\eu^{-\im \theta_1} \cos \omega - 1) \\
%     0     & \eu^{-\im \theta_2} \sin \omega  & \cdots & 0 & 
%                                        (\eu^{-\im \theta_2} \cos \omega - 1) \\
%     \vdots &&& 0 & \vdots \\
%     0 & 0 & \cdots & \eu^{-\im \theta_r} \sin \omega &  
%                                        (\eu^{-\im \theta_r} \cos \omega - 1) \\
%      (\eu^{-\im \theta_1} \cos \omega - 1) &
%      (\eu^{-\im \theta_2} \cos \omega - 1) &
%      \cdots & (\eu^{-\im \theta_r} \cos \omega - 1) &
%       -(\sin \omega \sum_{k=1}^r \eu^{-\im \theta_k} + c \omega)
%   \end{pmatrix}
% }
% \end{displaymath}
% for the variables $A_1, \dots, A_r, Z$. Note that formally the case $c=0$
% corresponds to the Kirchhoff boundary condition case and $M_0(\omega) =
% M(\omega)$. In a similar way as before we have
% \begin{displaymath}
%   \det M(\omega) = (\sin^{r-1}\omega) \eu^{-\im\sum_k \theta_k}
%      \bigl( 
%         2 \sum_k (\cos \omega - \cos \theta_k) - c \omega \sin \omega 
%      \bigr).
% \end{displaymath}
% Non-trivial solutions of the eigenvalue problem exist iff $\det M(\omega)=0$,
% i.e. if $\omega = \ell \pi$, $\ell \in \N$, or
\begin{equation}
  \label{eq:omega.theta.2}
  \cos \omega - \frac{\omega \sin \omega}{2rc} = 
      \frac 1 r \sum_{k=1}^r \cos \theta_k.
\end{equation}
Note that formally the case $c=0$ corresponds to the Kirchhoff boundary
condition case.  Again, the solutions $\omega = \ell \pi$ belong to Dirichlet
eigenfunctions on each edge and produce therefore bands degenerated to a point
$\{ (\ell\pi)^2 \}$.

\begin{thm}
  \label{thm:gaps.border}
  The limit operator $Q_0$ in the borderline case defined on a Cayley graph
  associated to an arbitrary finitely generated abelian discrete group
  $\Gamma$ has infinitely many spectral gaps located around
  $(2\ell+1)^2\pi^2/4$ provided $\ell$ is large enough. If at least one group
  order $p_j$ is odd, we have also spectral gaps below and above
  $(2\ell+1)^2\pi^2$ for $\ell=0,1,\dots$
\end{thm}
\begin{proof}
  The function $f(\omega):= \cos \omega - \omega \sin \omega /(2rc)$
  oscillates with an amplitude of order $\omega$. In particular, for
  $\omega=(2\ell+1)\pi/2$ we have $|f(\omega)|=(2\ell+1)\pi/(4rc)$, i.e.,
  there is no solution of~\eqref{eq:omega.theta.2} provided $\ell$ is large
  enough.  Furthermore, since $f((2\ell+1)\pi)=-1$, we can argue as in
  Theorem~\ref{thm:gaps.ex}.
\end{proof}

Returning to our graph-like periodic manifolds, we have the following
situation: In the case of fast decaying vertex volume, i.e.\ $(d-1)/d < \alpha
\le 1$, we have spectral gaps below and above $(2\ell+1)^2 \pi^2$ provided at
least one order $p_j$ is odd. In particular, $(2\ell+1)^2 \pi^2$ is an
isolated eigenvalue (degenerated band). In the borderline case, these gaps
remain open.  Furthermore, we always have spectral gaps around $(2\ell+1)^2
\pi^2/4$ provided $\ell$ is large enough. In the case of slowly decaying
vertex volume, i.e., $0 < \alpha < (d-1)/d$, all bands concentrate around
$\ell^2\pi^2$, i.e. we have large gaps around $(2\ell+1)^2 \pi^2/4$ for all
$\ell=0, 1, \dots$

The situation is more complicated if we allow different length ratios for the
edges. In such a case the spectrum could be more complicated; recall the
example of a lattice graph discussed in~\cite{exner-gawlista:96} shows where
number-theoretic properties of parameters play a r\^ole. This interesting
question will be considered separately.

%------------------------------------------------------------
\section*{Acknowledgements}
%------------------------------------------------------------
The authors appreciate P.~Kuchment who made available to them the paper
\cite{kuchment-zeng:03} prior to publication. O.P. is grateful for the
hospitality extended to him at the Nuclear Physics Institute of Czech Academy
of Sciences where a part of this work was done. The research was partially
supported by the GAAS grant A1048101 and the SPECT program of the European
Science Foundation.

%------------------------------------------------------------
% References
%------------------------------------------------------------

% Uncomment the next two lines for bibtex
%\bibliographystyle{amsalpha}% with numbers: amsplain, with letters: amsalpha
%\bibliography{/home/post/Mathematik/Literatur/literatur}

%--------------------------------
% include file "muster.bbl"
%--------------------------------

\providecommand{\bysame}{\leavevmode\hbox to3em{\hrulefill}\thinspace}
\newcommand{\etalchar}[1]{$^{#1}$} \def\cprime{$'$}

\end{document}